\documentclass[prb,showpacs,preprintnumbers,amsmath,amssymb,printfigures,twocolumn,superscriptaddress]{revtex4-1}
\usepackage{graphicx}
\usepackage{amsmath}
\usepackage{bm}
\usepackage[T1]{fontenc}
\usepackage{lmodern}
 \usepackage{textcomp}
\usepackage{amstext}
\usepackage[Symbol]{upgreek}
\usepackage{subfigure}
\usepackage{amsmath}
\usepackage{booktabs}
\usepackage{dcolumn}
\usepackage{color}
\usepackage{enumerate}
\usepackage{latexsym,bm}
\usepackage{longtable}
\usepackage[colorlinks=true, allcolors=blue]{hyperref}
\makeatletter
\begin{document}

\preprint{manuscript}
\title{First-principles Study on Structural, Thermal, Mechanical and Dynamic Stability of T'-MoS$_2$}
\author{Y. C. Liu}
\thanks{liuyachao@xaut.edu.cn (Y. C. Liu).}
\affiliation{Department of Applied Physics, Xi'an Jiaotong University, Xi'an 710049, China}
\affiliation{Department of Applied Physics, Xi'an University of Technology, Xi'an 710054, China}

\author{V. Wang}
\affiliation{Department of Applied Physics, Xi'an University of Technology, Xi'an 710054, China}

 \author{M. G. Xia}
\affiliation{Department of Applied Physics, Xi'an Jiaotong University, Xi'an 710049, China}
\author{S. L. Zhang}
\thanks{zhangsl@mail.xjtu.edu.cn (S. L. Zhang).}
\affiliation{Department of Applied Physics, Xi'an Jiaotong University, Xi'an 710049, China}

 \date{\today}

\begin{abstract}
Using first-principles density functional theory calculations, we investigate the structure, stability, optical modes and electronic band gap of a distorted tetragonal MoS$_2$ monolayer (T'-MoS$_2$). Our simulated scanning tunnel microscopy (STM) images of T'-MoS$_2$ are dramatically similar with those STM images which were identified as K$_{x}$(H$_{2}$O)$_{y}$MoS$_{2}$ from a previous experimental study. This similarity suggests that T'-MoS$_2$ might have already been observed in experiment but was unexpectedly misidentified. Furthermore, we verify the stability of T'-MoS$_2$ from thermal, mechanical and dynamic aspects, by \emph{ab initio} molecular dynamics simulation, elastic constants evaluation and phonon band structure calculation based on density functional perturbation theory,  respectively. In addition, we calculate the eigenfrequencies and eigenvectors of the optical modes of T'-MoS$_2$ at $\Gamma$ point and distinguish their Raman and infrared activity by pointing out their irreducible representations using group theory; at the same time, we compare the Raman modes of  T'-MoS$_2$ with those of H-MoS$_2$ and T-MoS$_2$. Our results provide a useful guidance for further experimental identification and characterization of T'-MoS$_2$.
\end{abstract}
\keywords{distorted tetragonal MoS$_2$; simulated STM images; first-principles; physical stability; Raman and infrared modes; band gap}
\pacs{ 68.37.Ef, 63.20.dk, 63.22.-m, 78.20.Ek}
\maketitle
\section{Introduction}
The successful fabrication of two-dimensional (2D) materials such as graphene have aroused intense interest due to their intriguing electronic, mechanical, optical, and thermal properties.\cite{Novoselov2005, Zhang2005} The gapless nature of graphene, however, presents limitations to their potential application in industry.\cite{Liao2010, Schwierz2010} Therefore, the interest of study has gradually turned to other 2D material such as transition metal dichalcogenides (TMD) monolayer. TMD generally have diverse crystal structures which can provide significantly different electronic properties varying from semiconducting to metallic.\cite{Mak2010, Wu2011, Kan2014, Li2014Gapless}

The most common TMD is MoS$_{2}$, which has three possible phases. They are H-MoS$_{2}$\cite{Wang2012},T-MoS2\cite{py1983structural, Wypych1998, ataca2012stable} and T'-MoS$_{2}$,\cite{py1983structural, Kan2014} as displayed in Fig.1 respectively. H-MoS$_{2}$, i.e., the 2D trigonal prismatic phase, being the most stable configuration under normal conditions,\cite{benavente2002intercalation} can be exfoliated from the bulk 2H phase (P6/mmc) using a mechanical method\cite{Wang2012} or be synthesized with vapor deposition\cite{Lee2012} It is a semiconductor with a direct band gap of 1.8 eV.\cite{Qin1991,Radisavljevic2011} T-MoS$_{2}$ (tetragonal symmetry, octahedral coordination) phase can be synthesized from solvent based exfoliation method. It is reported to be metallic and can be used as an electrode material.\cite{Kappera2014} Although T-MoS$_{2}$  was observed in experiment,\cite{lin2013atomic,Kappera2014} the stability of  T-MoS$_{2}$ is still a controversial issue. For example, density functional theory (DFT) calculations predict that the free standing T-MoS2 is unstable, since the seriously imaginary frequency presented in its phonon dispersion relation.\cite{Shirodkar2014emergence, Singh2015}

Qin et al. performed an STM study on the surface of restacked MoS$_{2}$ and observed a new superstructure characterized by the formation of zigzag chains.\cite{Qin1992} Then, an electron crystallography study also suggested that the restacked MoS$_{2}$  is more like WTe$_{2}$  with zigzag Mo-Mo chains.\cite{heising1999structure} This zigzag phase is called the distorted tetragonal MoS$_2$, labeled as T'-MoS$_{2}$ in the present paper, which was also referred to as the 1T' phase or ZT-MoS$_{2}$ phase in other theoretical study.\cite{py1983structural, Kan2014} T'-MoS$_{2}$  is thought to be a charge density wave (CDW) state as a result of the Piers phase transition from T phase.\cite{Whangbo1992} The structural stability of T'-MoS$_{2}$ was first inferred from formation energy by Kan et al..\cite{Kan2014} They found that the formation energy of T'-MoS$_{2}$ is higher than that of  H-MoS$_{2}$ but lower than T-MoS$_{2}$. Namely, T'-MoS$_{2}$ is a meta-stable phase. Qian et al. calculated the phonon band structure of T'-MoS$_{2}$ and found no imaginary frequency, which confirmed the vibrational stability of T'-MoS$_{2}$.\cite{Qian2014} As to the stability of T'-MoS$_{2}$ in other physical respects, such as the thermal and mechanical stability, however, has not been studied theoretically by far, to the best of our knowledge. Although previous experimental study has reported the observation of T'-MoS$_{2}$ phase identified by experimental STM images,\cite{Eda2012, Guo2015} those STM images, however, have not been sufficiently explicit to demonstrate the existence of T'-MoS$_{2}$. In addition, a discrepancy is also presented with respect to the band gap of T'-MoS$_{2}$. For instance, T'-MoS$_{2}$  is first predicted to be a semiconductor with a narrow band gap.\cite{Kan2014, Qian2014} In contrast, the electronic band structure given by Gao et al. implied that T'-MoS$_{2}$ was a semimetal.\cite{gao2015charge}

Therefore, a theoretical study on the physical stability and electronic properties of T'-MoS$_{2}$ is necessary and urgent. In the present work, we perform density functional theory (DFT) calculations within local density approximation (LDA) to investigate the simulated STM images, stability and the electronic band gap of T'-MoS$_{2}$. The simulated STM images provide a significant reference for identifying the lattice structure from experimental STM images. The \emph{ab initio} molecular dynamics (AIMD) simulations confirm the thermodynamic stability of T'-MoS$_{2}$ at room temperature; the calculating results of elastic constants meet the Born-Huang criteria, which implying the mechanical stability of T'-MoS$_2$; the absence of the imaginary frequency in the phonon dispersion relation indicates the vibrational stability of T'-MoS$_{2}$. Besides, we also classify the optical modes by group theory and compute their corresponding eigenfrequency and eigenvector, which play an important role in the identification and characterization of T'-MoS$_{2}$ phase from optical experiment. Moreover, we make a contrast calculation of the electronic band structure to determine the effect of the spin-orbit coupling, which clarifies the origin of band gap of T'-MoS$_{2}$.

The remainder of this paper is organized as follows. In Sec. II, methodology and computational details are described. Sec. III presents first the simulation of STM imaging of MoS$_2$ in three different phases, then the stability of  T'-MoS$_{2}$ is explored from different aspects. The symmetry classification of the vibrational modes along with their eigenfrequency and eigenvector are calculated. Furthermore, the electronic band structure and band gap of T'-MoS$_{2}$ are investigated by considering the spin-orbit coupling. Finally, conclusions are drawn in Sec. IV.
 \begin{figure}[htbp]
  \centering
  \includegraphics[scale=0.38]{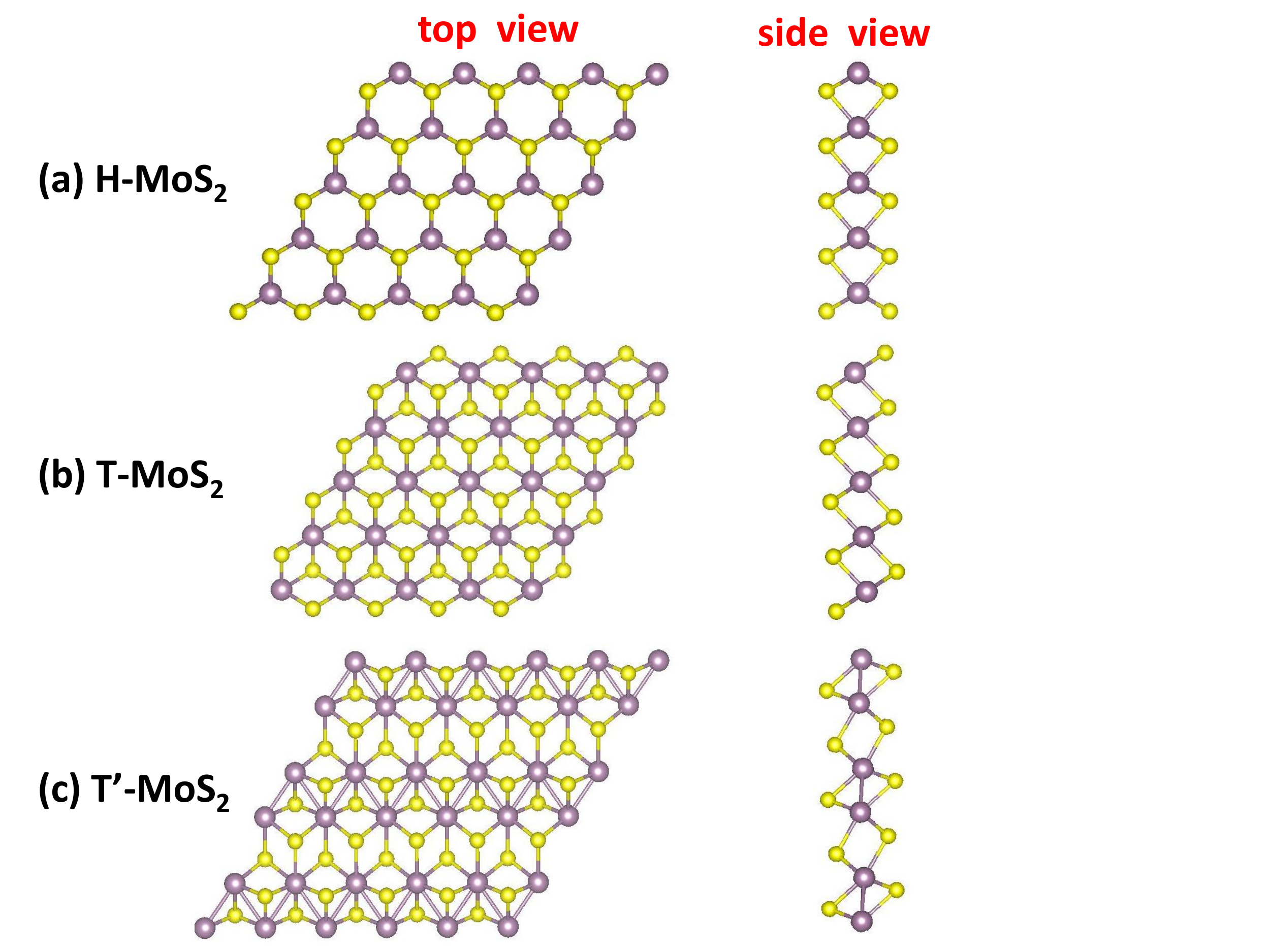}
  \caption{\label{3phases2}(Color online) Top view and side view of atomic structures of monolayer MoS$_2$ in H, T, and T' phases are shown in (a), (b), and (c), respectively.}
\end{figure}
\section{Methodology}
Both total energy and electronic band structure calculations were performed by using the Vienna ab initio simulation package (VASP).\cite{Kresse1996, Kresse1996a} The electron-ion interaction was described by using the frozen-core projector augmented wave (PAW) method;\cite{PAW, Kresse1999} the exchange and correlation were treated with generalized gradient approximation (GGA) in the Perdew-Burke-Ernzerhof (PBE) form.\cite{Perdew1996} Besides standard DFT with GGA, the hybrid Heyd-
Scuseria-Ernzerhof (HSE)06 method,\cite{Heyd2003Hybrid, Heyd2006Erratum} was also employed to give a more accurate description of the band gap of T'-MoS$_{2}$. In the HSE06 method, a fraction of the exact screened Hartree-Fock (HF) exchange is incorporated into the PBE exchange using a mixing parameter $\alpha$= 0.25. We used a cutoff energy of 300 eV for the plane wave basis set, which yields total energies convergence better than 1 meV/atom. The Van der Waals interactions are treated with the empirical correction scheme of Grimme's DFT-D2 method, which has been successful in describing the geometries of various layered materials.\cite{Grimme2006, Bucko2010}

In the slab model of single-layer MoS$_2$, the periodic slabs were separated by a vacuum layer of 15 {\AA} in the \emph{c} direction to avoid mirror interactions. A 10\texttimes{}5\texttimes{}1 \emph{k}-mesh including $\Gamma$-point, generated according to the Monkhorst-Pack scheme,\cite{Monkhorst1976} was applied to the Brillouin-zone (BZ) integrations. Through geometry optimization, both the shapes and internal structural parameters of pristine unit-cells were fully relaxed until the residual force on each atom is less than 0.01 eV/{\AA}.

To examine the stability of the modeled structure of T'-MoS$_{2}$ from the lattice dynamics point of view, the force-constant approach involving a finite displacement was adopted as employed by the \href{http://atztogo.github.io/phonopy/index.html}{PHONOPY}\cite{phonopy} code. The force constant matrix was calculated with a 7\texttimes{}4\texttimes{}1 supercell containing 168 atoms, based on the density functional perturbation theory (DFPT) method \cite{Gonze1997} implemented into VASP. Furthermore, the phonopy codes also enable us to obtain the eigenfrequency and eigenvector of lattice vibrational modes at the center of BZ.

The simulated STM images were generated by using the \href{http://www.p4vasp.at/}{P4VASP} package, which can facilitate the simulation of STM image with a continuously varying scanning distance. The theory for simulating  STM imaging by {\em ab initio} density functional calculations is well established.\cite{Tomanek1988} Giving a small bias voltage $V_{b}$ between the sample and the STM tip produces a tunneling current, whose density $j({\bf r})$ can be obtained from a simple extension \cite{Selloni1985} of the expression derived by Tersoff and Hamann \cite{Tersoff1983, Tersoff1985}:

 \begin{equation}
j({\bf r},V_{b}) {\propto} \rho_{\bf STM}({\bf r},V_{b}),
\end{equation}
where
\begin{equation}
\rho_{\bf STM}({\bf r},V_{b})=\int_{E_F-eV_{b}}^{E_F}
\rho({ r},E) dE%
\label{eq2}
\end{equation}
and
\begin{equation}
\rho({\bf r},E)=\sum_{n,{\bf k}} |\psi_{n{\bf k}}({\bf r})|^2
                \delta(E_{n,{\bf k}}-E) \;.
\label{eq3}
\end{equation}
Here, $\rho({\bf r},E)$ is the local density of states at the center of the tip at ${\bf r}$ and $\psi_{n{\bf{k}}}({\bf r})$ are the electron eigenstates of the unperturbed surface at energy $E_{n,{\bf k}}$. These eigenstates are commonly represented by Kohn-Sham eigenstates obtained using DFT. The assumptions behind this is that the relevant tip states are described by $s$ waves with a constant density of states.\cite{Tersoff1983, Selloni1985, Tersoff1985} Furthermore, the tunneling matrix element is considered to be independent of both the lateral tip position for a constant tip-to-surface distance and the bias voltage $V_{b}$ in the narrow (but nonzero) energy region $[E_F-eV_{b},E_F]$. Equation \ref{eq3} describes tunneling from occupied states of the sample to the tip. The simulated STM image is not sensitive to the bias voltage as long as the valence band enters in the integral range, but sensitive to the scanning distance from the tip to the sample surface.

The simulating STM imaging has been used for studying the modification of the electronic structure of the 2H phase MoS$_2$ (0001) surface produced by several point defects.\cite{Fuhr2004} Recently, it was also used for exploring the few-layer phosphorus capped by graphene and hexagonal boron nitride monolayer.\cite{Rivero2015} In present work, we apply this method to study the structure of single-layered MoS$_2$. Different bias voltage $V_{b}$  are used for distinct phases of monolayer MoS$_2$ according to their electronic properties. For H-MoS$_2$, its band gap is 1.7 eV, and its Fermi level is under the conductor band at about 0.1 eV, so that the value of bias voltage $V_{b}$ is set to 1.8 V, and thus the energy range $[E_F-eV_{b},E_F]$ enters the valence band at about 0.2 eV. As to metallic T'-MoS$_2$, we have compared the simulated images using two different bias voltage (0.3 and 1.8 V) but find no significant distinction, hence we always use the smallest one in the following calculation. For T'-MoS$_2$, its band gap is merely 0.1 eV, so a bias voltage $V_{b}$=0.5 V is enough.
\section{Results and discussion}
\subsection{Simulated STM images and Identification of Monolayer MoS$_2$}
\begin{figure}[htbp]
           \centering
           \includegraphics[scale=0.42]{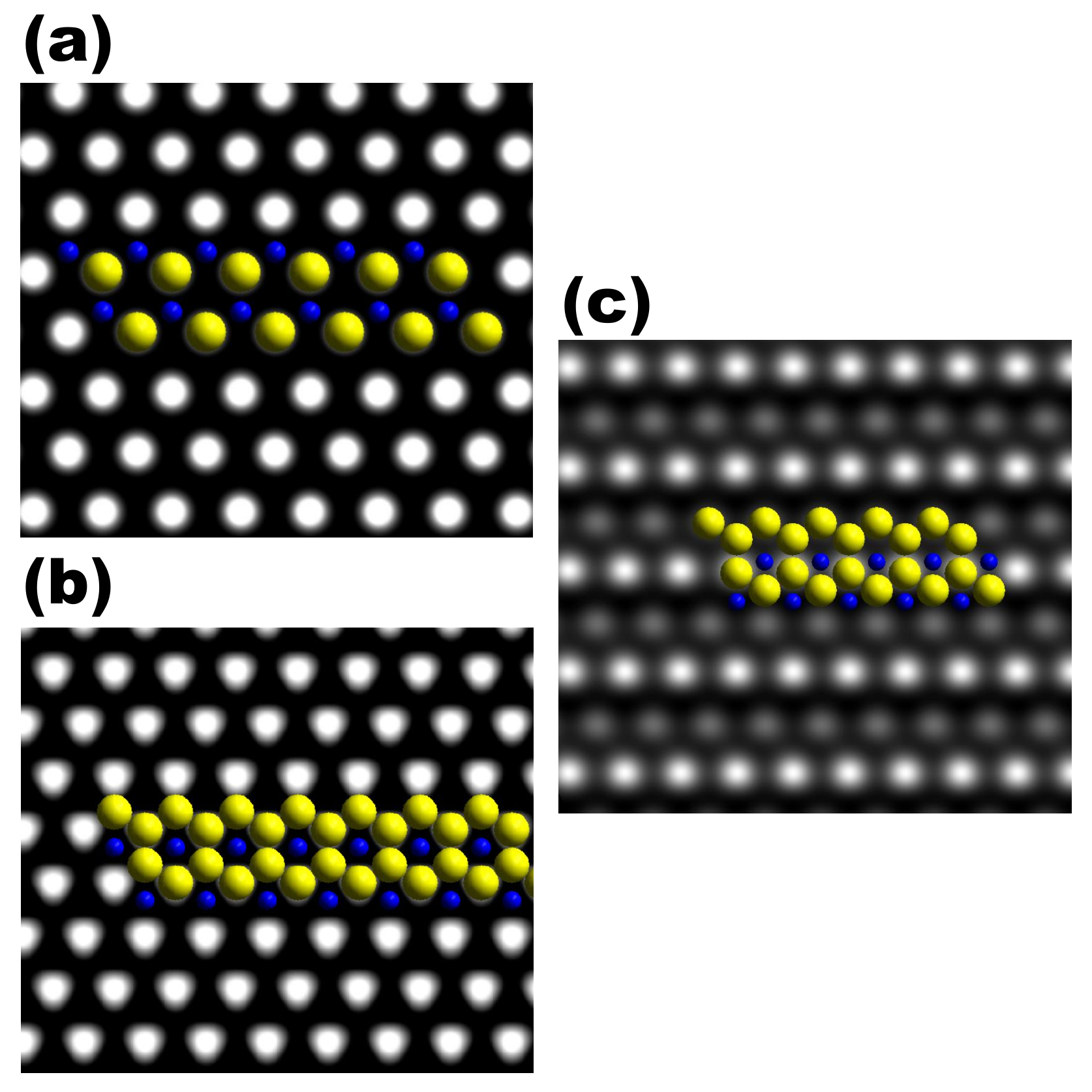}
           \caption{\label{STM-exp-sim}(Color online) Simulated STM images of monolayer MoS$_2$.  (a), (b), and (c) are simulated images of  H, T, and T' phases, respectively, where purple and yellow spheres represent Mo and S atoms.}
       \end{figure}

We begin our discussion by comparing the simulated STM images of the three possible structures, namely, H, T, and T' phases of monolayer MoS$_2$.\cite{chhowalla2013} The lattice structures of the three phases are displayed in  Fig. \ref{3phases2}. The most energetically favorable H-MoS$_2$ (as shown in Fig. \ref{3phases2}(a)) has a sandwich-like structure of three planes of 2D hexagonally packed atoms, S-Mo-S, where Mo atoms are trigonal-prismatically coordinated by six S atoms, forming ABA stacking with P6m2 space-group symmetry. In contrast, the Mo atoms in the T-MoS$_2$ (as shown in Fig. \ref{3phases2}(b)) structure are octahedrally coordinated with the nearby six S atoms, resulting in ABC stacking  with  P3m1 space group symmetry. H- and T-MoS$_2$ phase have very different electronic properties: the former is a large gap semiconductor but the latter a metal. It has been predicated that the T-MoS$_2$ is typically unstable in free-standing condition\cite{Shirodkar2014emergence, Kan2014}. T-MoS$_2$ should undergo the Piers distortion in one direction to form a $2 \times 1$ super-lattice structure, consisting of one-dimensional zigzag Mo-Mo chains along the other direction, \emph{i.e.}, the T'-MoS$_2$ phase, as shown in Fig. \ref{3phases2}(c). It implies theoretically that T'-MoS$_2$ should be more stable than T-MoS$_2$ in free-standing conditions. In experiments, however, Eda et al. have observed both T- and T'-MoS$_2$ by scanning transmission electron microscopy (STEM) imaging,\cite{Eda2012}  but the image of T'-MoS$_2$  is not so clear as that of T-MoS$_2$. Although WS$_2$ and MoTe$_2$ monolayer have been found experimentally.\cite{Mahler2014, Keum2015} T'-MoS$_2$, has not been identified unanimously in experiment yet, to the best of our knowledge. Therefore, we perform an \emph{ab initio} density functional calculations to simulate STM images of  MoS$_2$ monolayer in the three phases. Figure.\ref{STM-exp-sim} shows the calculated STM images of  H-, T-, as well as T'-MoS$_2$, respectively.  Our simulated STM images agree well with those images obtained in previous experiments.\cite{Eda2012} This agreement indicates the reliability of the simulated STM imaging method.

            \begin{figure}[htbp]
            \centering
            \includegraphics[scale=0.30]{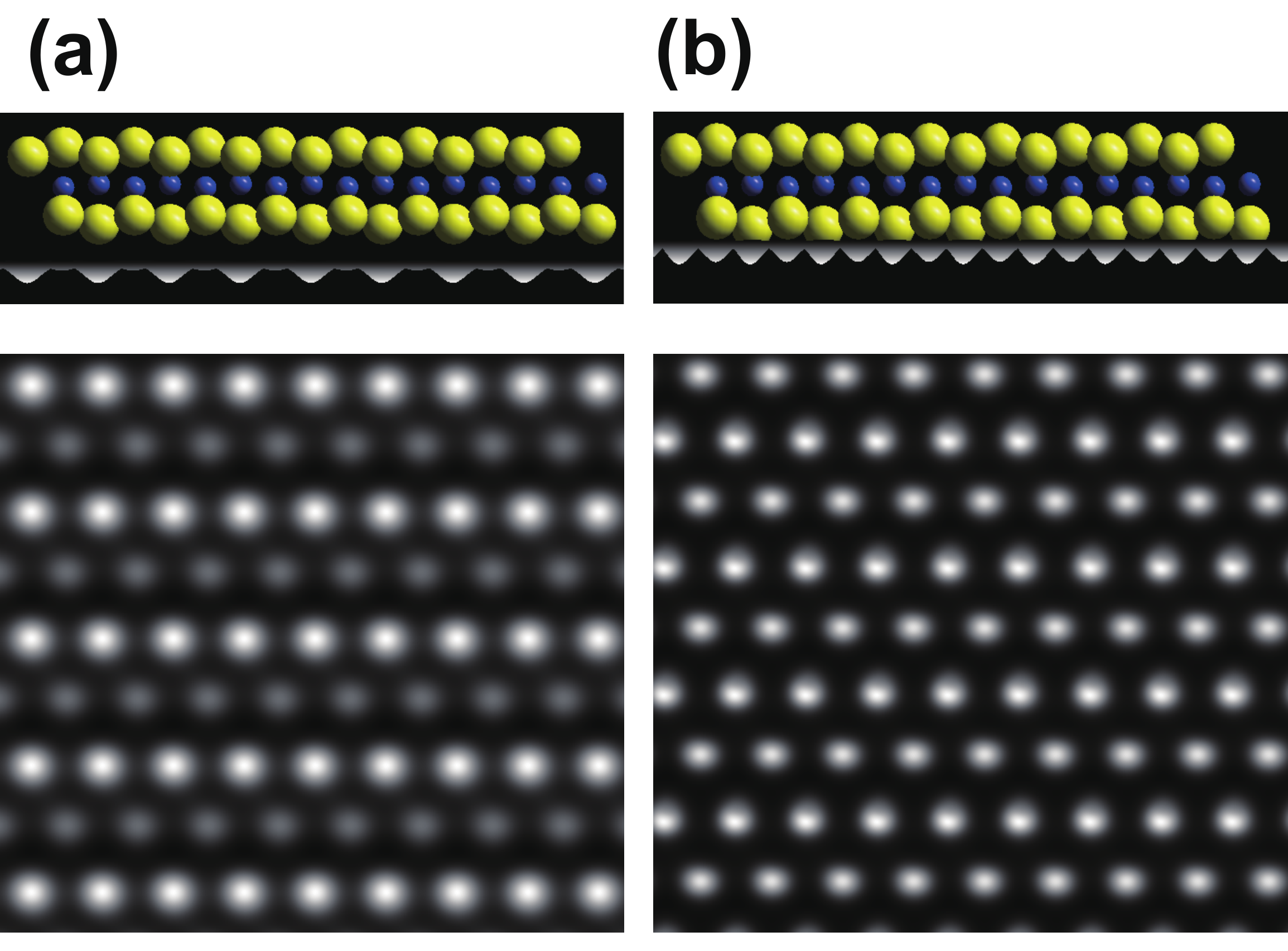}
              \caption{\label{STM-2T'}(Color online) Simulated STM images of T'-MoS$_2$ with different scanning distance. (a) the scanning distance $d=6.59$ ${\AA}$ and (b) the scanning distance $d=5.74$ ${\AA}$.}
          \end{figure}

\begin{figure}[htbp]
  \centering
  \includegraphics[scale=0.26]{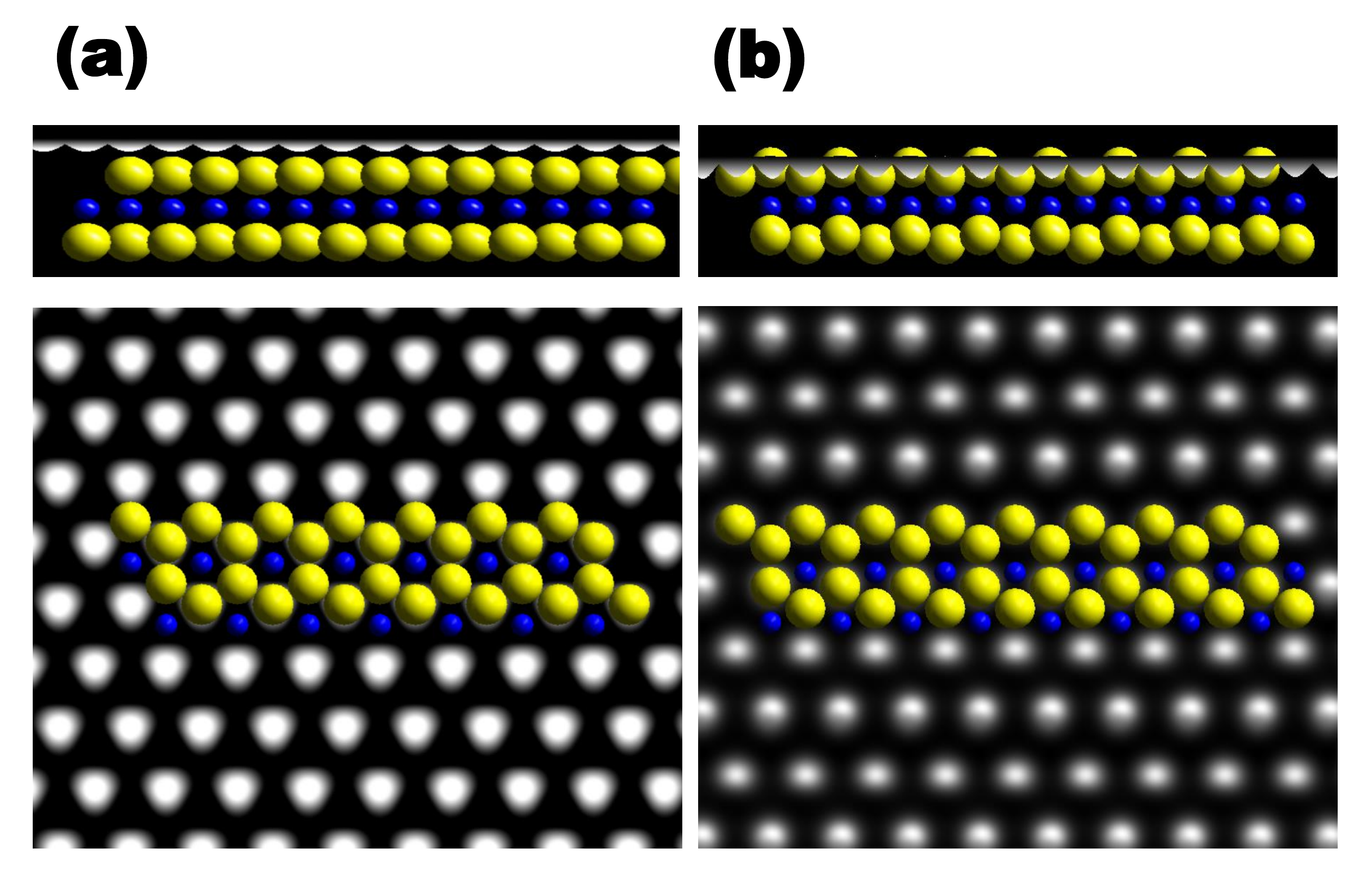}
         \caption{\label{STM-T-T'}(Color online) Simulated STM images of  MoS$_2$ with T, and T' phases. (a) and (b) are  T-MoS$_2$ and T'-MoS$_2$, respectively. For better interpretation of the images, we reproduced the atomic arrangement on the top of the images. On the higher sublets, the side view of the atomic arrangement of the monolayer and intensity profile are shown; on the lower sublets, the top view of those are given, respectively.}
          \end{figure}

The scanning distance $d$ represents the distance form scanning plane to referring plane, and the referring plane is put on the upper surface of crystal cell in slab model of MoS$_2$ monolayer. The scanning distance is denoted in term of the tip position in P4VASP. To determine the influence of the scanning distance on STM imaging, we perform the simulating  STM imaging calculations varying continuously with scanning distance. It is found that the simulated STM images vary remarkably with the scanning distance. This variation even may lead to misidentification of the experimental STM images.\cite{Altibelli1996} On one hand, it means that we may obtain quite different STM images actually belonging to the identical structure in experiment, as seen in Fig. \ref{STM-2T'}, in which we have shown the top and side view of two different scanning distances, while the tip position for the middle plane of the MoS$_2$ monolayer is $d=3.39$ ${\AA}$. Then the distances from scanning plane to the middle plane are $3.20$ ${\AA}$ and $2.35$ ${\AA}$, respectively. If merely judging from the top view of the simulation images without referring the other information, you must think that the lower sublet of Fig. \ref{STM-2T'}(a) represents T'-MoS$_2$  phase but that of Fig. \ref{STM-2T'}(b) belongs to the image of  T-MoS$_2$. On the other hand, it also means that the different phases of MoS$_2$ may have similar STM images. Comparing the simulated images of  T-MoS$_2$ and T'-MoS$_2$ with appropriate scanning distances presented in Fig. \ref{STM-T-T'}, you may find it is hard to distinguish these two phases. Thus, we should not make identification only by single experimental STM image without any other information.

Knowing this how can we identify the lattice structure of STM image in experiment? The method is to scan the STM images while varying with tip-to-surface distance. By comparing and contrasting those images, you can make the correct identification,  for the different structures have distinct changing patterns. This suggestion is deduced from our STM imaging simulation of T-MoS$_2$ and T'-MoS$_2$ with continuously varying scanning distance. The structural symmetry of simulated image of  T-MoS$_2$ remains invariant as the scanning distance varies consecutively, in contrast, that of  T'-MoS$_2$  is varying significantly, just as demonstrated in Fig.\ref{STM-2T'}. It is worth mentioning that the STM images obtained experimentally are usually scanning within one or two given tip-to-surface distance. Then a question arises: whether it is possible that the structure of MoS$_2$ observed in previous experiments could be T' phase rather than T phase?

\begin{figure}[htbp]
             \centering
            \includegraphics[scale=0.42]{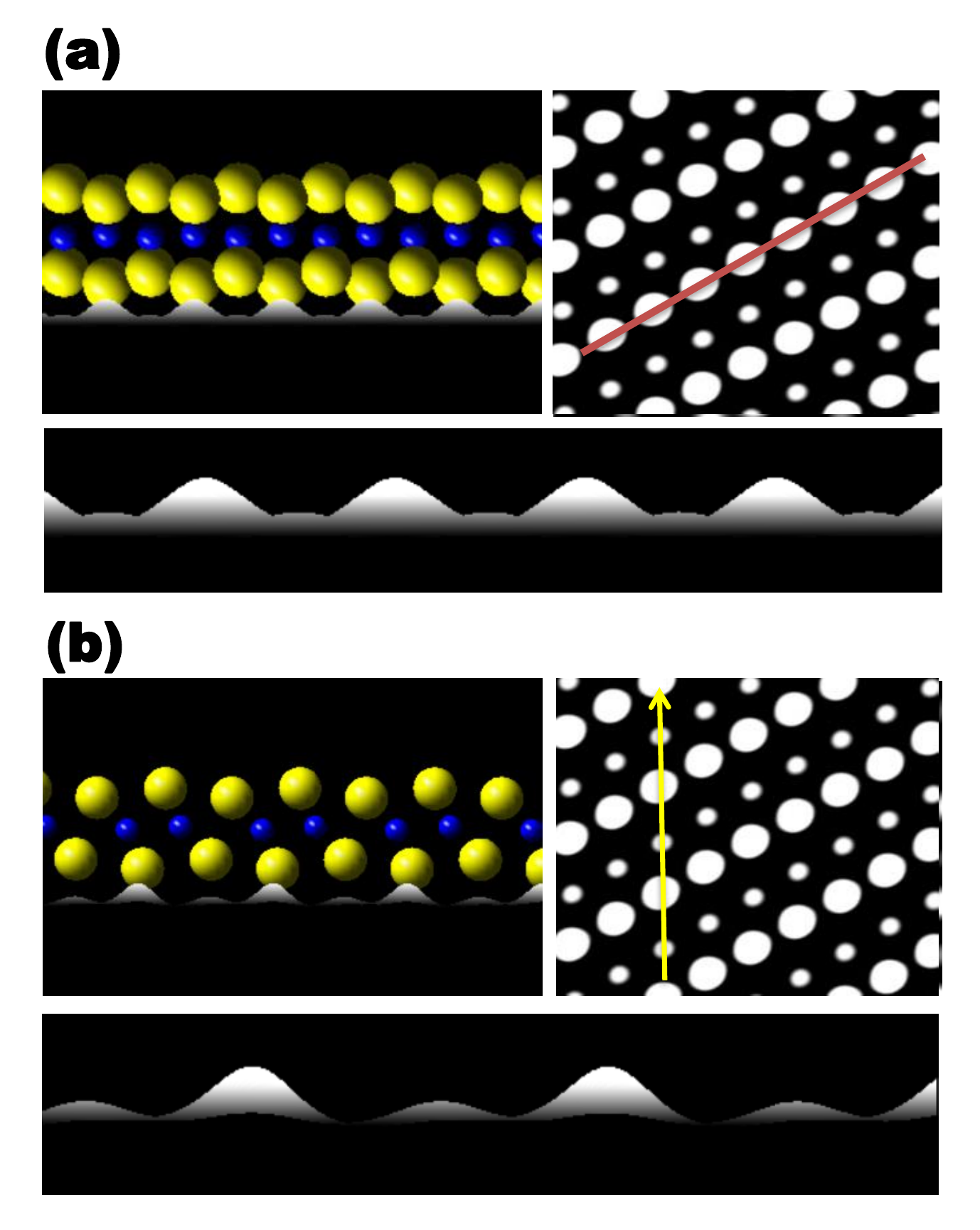}
            \caption{\label{exp-T'-sim-2}(Color online) (a) and (b) show the side view of atomic structure and intensity profiles along the lines indicated in simulated STM images of T'-MoS$_2$, respectively.}
 \end{figure}

It is natural to examine the relevant experiments in literature, and we find that the answer is affirmative. In an experimental STM imaging study of  T-MoS$_2$\cite{Wypych1998}, there are several STM images in Figures 2 and 3 in Ref[\onlinecite{Wypych1998}], which were identified as K$_{x}$(H$_{2}$O)$_{y}$MoS$_{2}(x<0.3)$. We make the corresponding simulation of T'-MoS$_2$, which are shown in Fig. \ref{exp-T'-sim-2}. Compare our simulated STM images of T'-MoS$_2$ phase with these images, we find that the simulated images surprisingly accord with the experimental STM images. This dramatic accordance indicates strongly that these experimental images should be corresponding to T'-MoS$_2$ phase rather than K$_{x}$(H$_{2}$O)$_{y}$MoS$_{2}$. That is to say, the T'-MoS$_2$ has been synthesized accidentally but misidentified unfortunately. If this was true, it actually means an experimentally feasible method for synthesizing T'-MoS$_2$, which is important for the fabrication of a novel topological field effect transistor.\cite{Qian2014}

The instability of free standing T-MoS2 at 0 K is revealed by imaginary frequency presented in its phonon dispersion relations from the recent first-principles calculations.\cite{Shirodkar2014emergence, Singh2015} At the same time, several theoretical and experimental researches show that the function of alkali metal is to offer an extra electron to make the T-MoS$_2$ phase more stable in energy.\cite{Kappera2014} While in Ref. [\onlinecite{Wypych1998}], the presence of water stabilizes the T' phase from the original high symmetric structure.\cite{Qin1991, Yang1991, Qin1992, Gordon2002}. Therefore, the method designed to obtain T-MoS$_2$ is actually a feasible method to produce T'-MoS$_2$ in experiment. For reliably identifying T'-MoS$_2$ in experiment, it is necessary to exploit the otherwise stability of  T'-MoS$_2$.

\subsection{Thermal stability of T'-MoS$_2$}
The thermal stability of T'-MoS$_2$ is explored by performing AIMD simulations using canonical ensemble. To reduce the constraint of periodic boundary condition, the T'-MoS$_2$ is simulated by ($3\times2$) super-cells. The snapshots of T'-MoS$_2$ atomic configurations for the final stages of AIMD simulations at 300 K and 800 K are shown in Fig. 6 (a) and (b), respectively. One can find that no significantly reconstruction are observed at 300 K and 800 K. Here the only exception in the latter case is that the S and Mo atoms are found to be slightly moved due to thermal fluctuation. This means that T'-MoS$_2$ can withstand the higher temperature at least up to 800 K, implying the high-energy barriers between T' phase and H phase, which is in consistent with the first-principles calculations performed by Qian et al.\cite{Qian2014}
 \begin{figure}[htbp]
\centering
\includegraphics[scale=0.16]{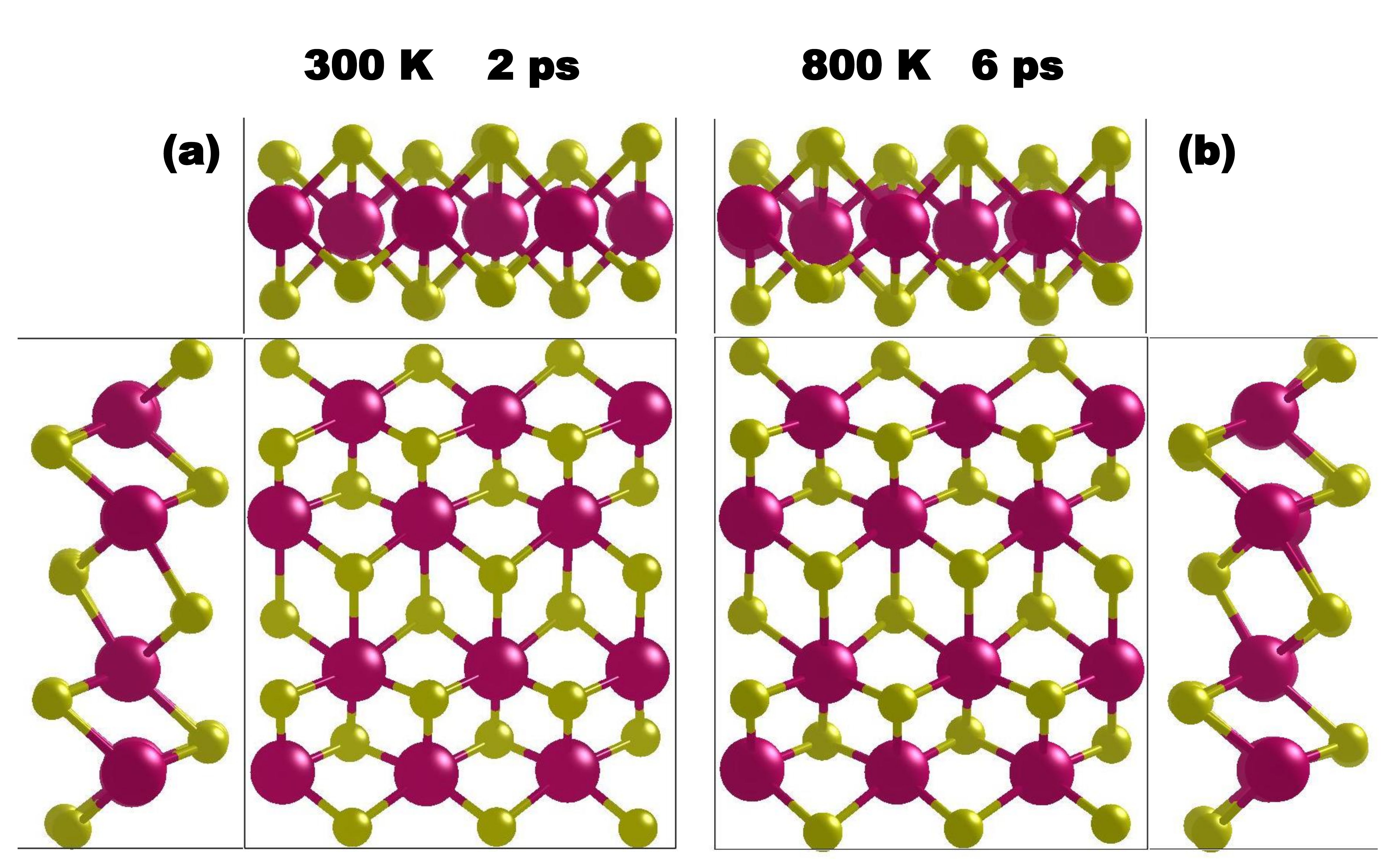}
\caption{\label{MD-simulation}(Color online) Snapshots of atomic configurations of T'-MoS$_2$ at the end of AIMD simulations from front, top, and side views, respectively. The simulated super-cells are marked by black squares, and their corresponding temperature and time are denoted above each panel.}
\end{figure}
\subsection{Mechanical Stability and Anisotropy of T'-MoS$_2$}
Since the super-cell is fixed during the MD simulations, we have to evaluate the effect of elastic distortion on structural stability. In order to guarantee the positive-definiteness of strain energy following lattice distortion, the components of linear elastic modulus tensor of a stable crystal must obey the Born-Huang criteria \cite{Ding2013}. We calculate the change of energy due to the in-plane strain to examine the mechanical stability of T'-MoS$_2$. For a 2D crystal, the elastic strain energy per unit area can be written as\cite{Wang2015}
 \begin{align}
 U(\varepsilon)=\frac{1}{2}C_{11}\varepsilon_{xx}^{2}+\frac{1}{2}C_{22}\varepsilon_{yy}^{2}+ C_{12}\varepsilon_{xx}\varepsilon_{yy}+2C_{66}\varepsilon_{xy}^{2},
  \end{align}
where $C_{ij}$ are the components of the elastic modulus tensor using the standard Voigt notation ( i.e., 1-xx, 2-yy, and 6-xy),\cite{PhysRevB.85.125428} corresponding to second partial derivative of elastic energy with respect to strain. The elastic constants can be derived by fitting the energy curves associated with uniaxial and equi-biaxial strains. Under uniaxial strain applied along x direction, $\varepsilon_{yy}=0$, this leads to $U(\varepsilon)=\frac{1}{2}C_{11}\varepsilon_{xx}^{2}$. Parabolic fitting of the uniaxial strain curve yields $C_{11}=109.0$ GPa$\cdot$nm. Similarly, under uniaxial strain applied
along y direction, $C_{22}$ is derived to be 124.0 GPa$\cdot$nm. Under equi-biaxial strain, $\varepsilon_{xx}=\varepsilon_{yy}$, one have $U(\varepsilon)=(\frac{1}{2}C_{11}+\frac{1}{2}C_{11}+C_{12})\varepsilon_{xx}^{2}$. By fitting the equi-biaxial strain curve, we obtain $(\frac{1}{2}C_{11}+\frac{1}{2}C_{11}+C_{12})=130.9$ GPa$\cdot$nm, which means that $C_{12}=14.4$ GPa$\cdot$nm. For a mechanically stable 2D crystal, the elastic constants should satisfy two criteria: $C_{11}C_{22}-C_{12}>0$ and $C_{66}>0$.\cite{Ding2013}For T'-MoS$_{2}$, one can easily verify that the calculated components of the elastic modulus tensor satisfy $C_{11}C_{22}-C_{12}>0$; besides, the calculated $C_{66}$=38.8 GPa$\cdot$nm is positive. Both the two criteria of mechanical stability are met, thus the mechanical stability of T'-MoS$_{2}$ is confirmed.

Due to its lower point group symmetry, T'-MoS$_{2}$ has anisotropic elastic property, which is significantly different from H- and T-MoS$_{2}$. Both H- and T-MoS$_{2}$ are of isotropic elastic properties described by two elastic constants: Young's modulus $Y$ and Poisson's ratio $\nu$. The Young's modulus and Poisson's ratio of T'-MoS$_{2}$, however, do not remain constant, but vary with orientation. The formula for Young's modulus $Y(\theta)$ and Poisson's ratio $\nu(\theta)$ are\cite{Ding2013,wang2016lattice}
 \begin{align}
 \label{eq10}
Y(\theta ) = \frac{{{C_{11}}{C_{22}} - C_{12}^2}}{{{C_{11}}{s^4} + {C_{22}}{c^4} + (\frac{{{C_{11}}{C_{22}} - C_{12}^2}}{{{C_{66}}}} - 2{C_{12}}){c^2}{s^2}}},\\
 \label{eq11}
\nu (\theta ) = \frac{{{C_{12}} + (\frac{{{C_{11}}{C_{22}} - C_{12}^2}}{{{C_{66}}}} - 2{C_{12}} - {C_{11}} - {C_{22}}){c^2}{s^2}}}{{{C_{11}}{s^4} + {C_{22}}{c^4} + (\frac{{{C_{11}}{C_{22}} - C_{12}^2}}{{{C_{66}}}} - 2{C_{12}}){c^2}{s^2}}},
 \end{align}
where $s = sin(\theta)$ and $c = cos(\theta)$, $\theta$ is the the angle with respect to the x-axis. The above two formulas are universal for all orthogonal 2D crystal.

Their extremum directions can be determined by analyzing the zeros of the first derivative of $Y(\theta)$ and $\nu(\theta)$. Generally, there are three extremum directions for both $Y(\theta)$ and $\nu(\theta)$, in which there are two same extremum conditions: $sin(\theta)=0$ and $cos(\theta)=0$, which means that the coordinate axes directions x and y (rather, the symmetrical principal axes directions) are extremum directions. we find that for $Y(\theta)$ , both the two axes may maximum (or minimum) directions, meanwhile, for $\nu(\theta)$, they must be both minimum (or maximum) directions. Certainly, the two maximum (or minimum) directions mean that there must be one minimum (or maximum) direction between the two axes. Depending on the relative value of elastic constants, the third extremum between the two axes for $Y(\theta)$ may not exist, then one axis direction is maximum and the other minimum direction. For Poisson's ratio, the third extremum always exists, except for isotropic case. The above conclusions about extremum direction are also universal for orthogonal 2D crystal structures.

\begin{figure}[htbp]
  \centering
  \includegraphics[scale=0.34]{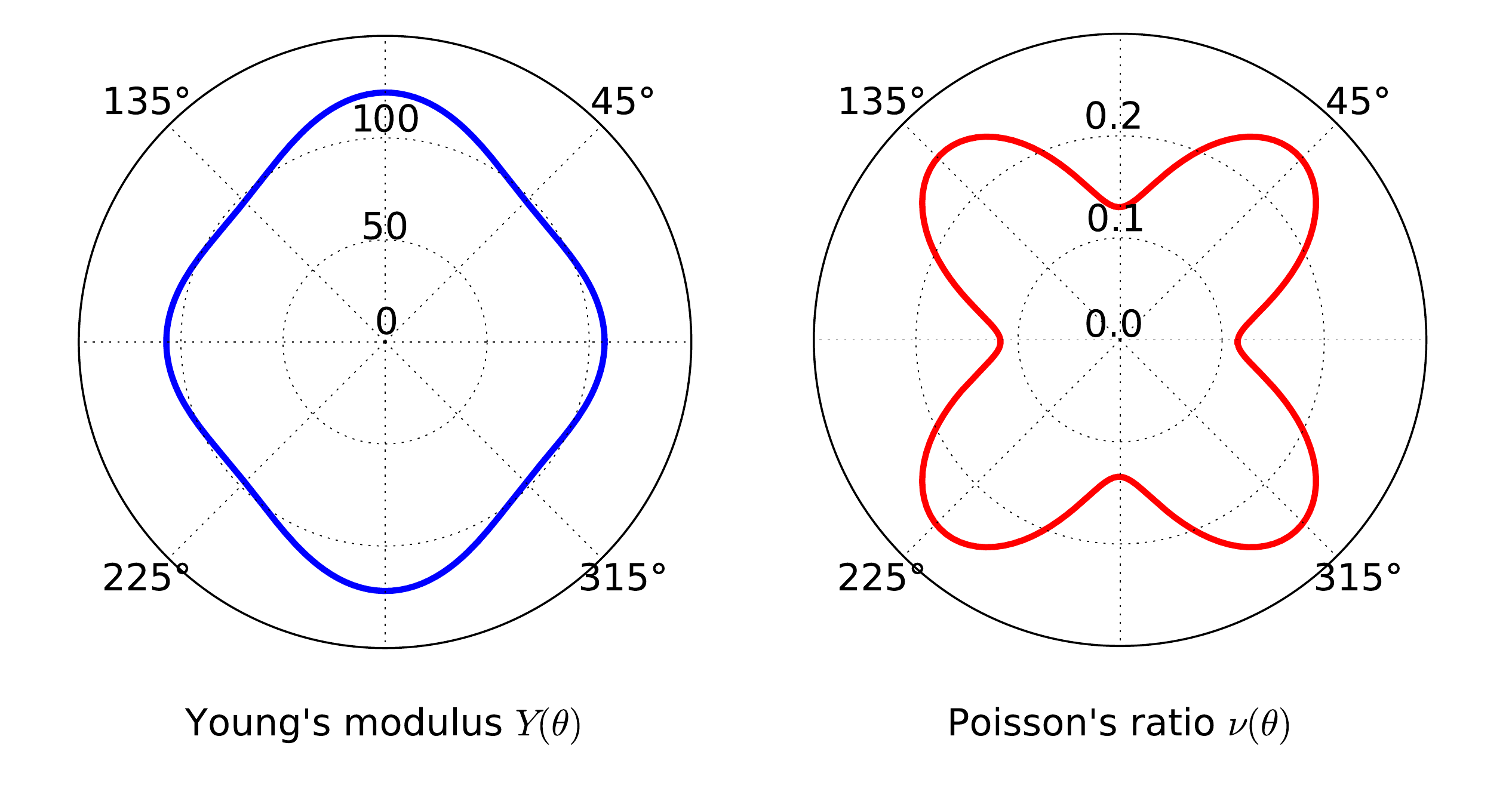}
  \caption{\label{Youngs-modulus}(Color online) Calculated orientation-dependent
 Young's modulus $Y(\theta)$ and Poisson's ratio $\nu(\theta)$ of T'-MoS$_2$.}
\end{figure}

 Here, we plot the curves of $Y(\theta)$ and $\nu(\theta)$ in polar coordinates for T'-MoS$_{2}$  in Fig \ref{Youngs-modulus}, which intuitively show the elastic anisotropy of T'-MoS$_{2}$. First, it can be seen that the extremum directions are consistent with our analyses above. Second, it can be found that the variation range of Young's modulus is from about 96.9 to 124.0 GPa$\cdot$nm, the value of Poisson's ratio is limited between 0.117 and 0.25. Third, it can be found that the principal minimum direction of $Y (\theta)$ is along $\theta=39.5^{\circ}$ and the maximum direction of $\nu(\theta)$ for T'-MoS$_{2}$  are almost along diagonal direction.
\subsection{Lattice Dynamic Stability and Vibrational Modes of T'-MoS$_{2}$}
\subsubsection{Lattice dynamic stability}
\begin{figure}[htbp]
\centering
\includegraphics[scale=0.42]{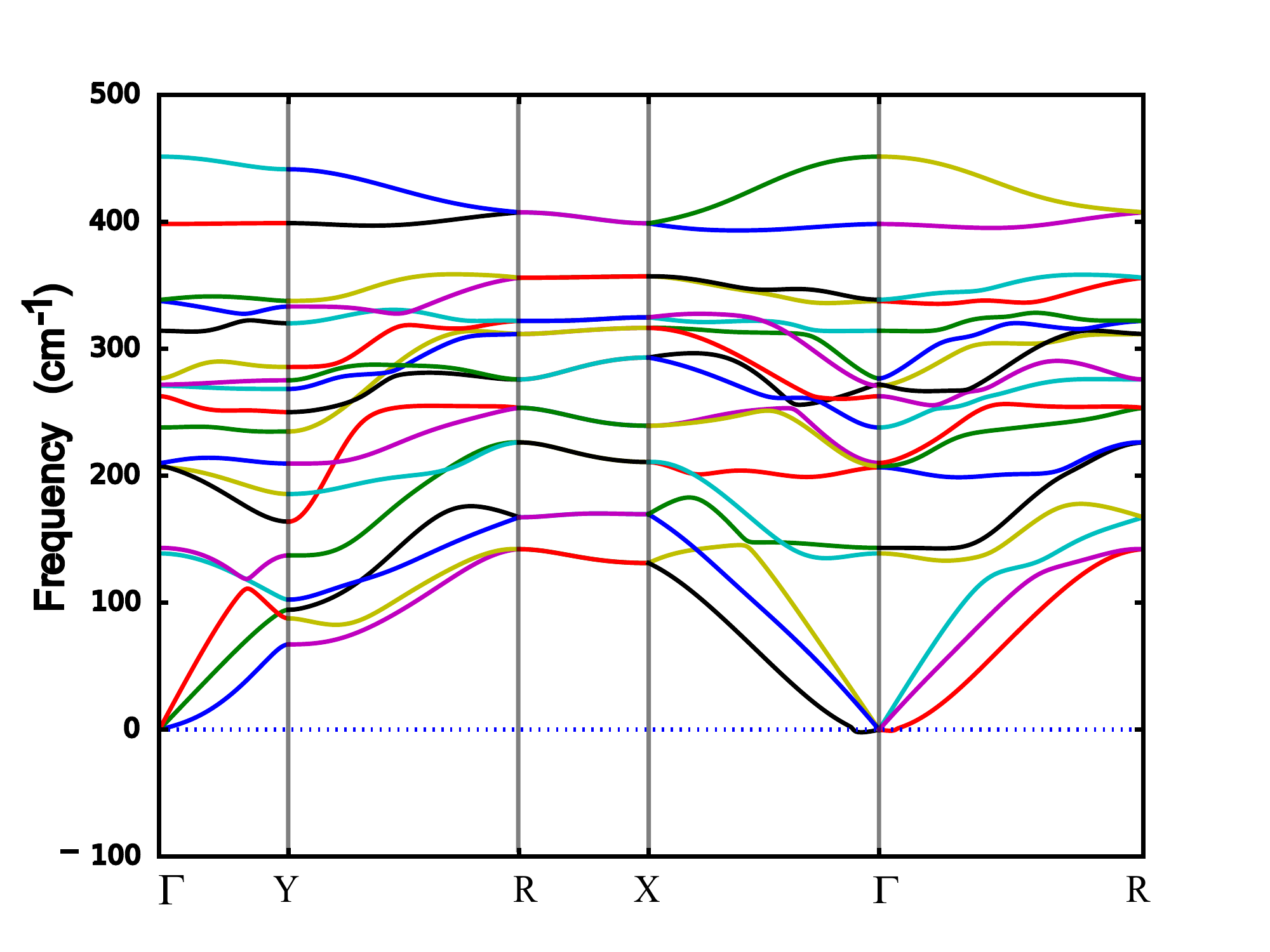}
\caption{\label{phonon-band}(Color online) The phonon band structure of T'-MoS$_2$ displaying the connection of vibrational bands. The absence of imaginary frequency indicates the stability of T-MoS$_2$.}
\end{figure}

To further verify the lattice dynamic stability of the T'-MoS$_2$, we calculate phonon dispersion relation of  T'-MoS$_2$ and demonstrate it in Fig. \ref{phonon-band}. we can note that the phonon dispersion of T'-MoS$_2$ has three acoustic and fifteen optical branches. The three acoustic branches are the in-plane longitudinal acoustic (LA), the transverse acoustic (TA), and the out-of-plane acoustic (ZA) branches. The LA and TA branches have linear dispersion and a higher frequency than the ZA mode around $\Gamma$ point in the Brillouin zone. In contrast to H-MoS$_2$,\cite{JimenezSandoval1991, Molina-Sanchez2011} there is no band gap between acoustic branches and optical branches. All the rest of vibrational branches along other lines in BZ are non-degeneracy, except the vibrational branches along R--X line at the boundary of BZ, which are two order degeneracy. The lifting of degeneracy of vibrational branches reveals the Piers phase transition from a high symmetric structure. The absence of the imaginary frequency throughout the 2D BZ indicates the structural stability of the T'-MoS$_2$. Our results are in good agreement with those obtained in Ref [\onlinecite{ Qian2014}].

\subsubsection{Symmetric analysis of lattice vibrational modes}

Both laser Raman scattering and Infrared absorption spectra are powerful tools for structural identification and characterization of 2D materials. To guide the optical spectra study in future experiment, we deduce the symmetry classification of  phonon modes at the $\Gamma$ point by using group theory, and further point out R and IR activity of the optical modes. The unit cell of T'-MoS$_2$ consists of two S-Mo-S units with a total of six atoms, suggesting that there are eighteen phonon modes (three acoustic and fifteen optical modes) at the $\Gamma$ point. Lattice vibrations can be classified based on the irreducible representation of space group.\cite{Dresselhaus2007group} The space group of T'-MoS$_2$ is $C^{2}_{2h}$ (or $P2_{1}/m$, No.11), whose factor group is isomorphic with the point group $C_{2h} $. The character table for point group $C_{2h} $  is given in Table \ref{character},
\begin{table}[htbp]
\centering
 \begin{ruledtabular}
\caption{\label{character}Character table for the point group $C_{2h}$ including basis functions of the irreducible representations.}
\begin{tabular}{ccccccc}
   $\rm{SG}$&$\rm{PG}$&$E$ &$ C_{2}$ &$\sigma$&$ i$&$\rm{basis}$\\
    \hline
   $\Gamma_{1}^{+}$  &$A_{g}$ &1&{ 1}&{ 1}&{ 1}&$R_{z},x^{2},y^{2},z^{2},xy$\\
   $\Gamma_{1}^{-}$  &$B_{g}$  &1&-1&-1&{ 1}& $R_{x},R_{y},xz,yz$\\
   $\Gamma_{2}^{+}$ &$A_{u}$  &1&{ 1}&-1&-1&$z$ \\
   $\Gamma_{2}^{-}$  &$B_{u}$  &1&-1&{ 1}&-1&$x,y$\\
   \end{tabular}
 \end{ruledtabular}
\end{table}
where $A_{g}$, $A_{u}$, $B_{g}$ and $B_{u}$ are signs of one-dimensional irreducible representations; $A$ and $B$ are used when the character of the major rotation operation is 1 or -1, respectively; the subscripts $ g (gerade)$ and $ u (ungerade)$ denote representations that are symmetric and antisymmetric with respect to the inversion operation if the point group has a center of inversion symmetry; $x$, $y$, and $z$ are components of polar vectors. From Table \ref{character}, we note that T'-MoS$_2$ has no two-dimensional irreducible representations, i.e., there is no degenerate optical modes at the center of BZ, which distinctively differs from that of H- and T-MoS$_2$.\cite{Wieting1971, JimenezSandoval1991, Cai2014, Zhang2015}
\begin{table}[htbp]
\centering
\begin{ruledtabular}
\caption{\label{Reduced Representation} Characters of  vector, equivalent, and vibration representations for T'-MoS$_2$. }
\begin{tabular}{ccccc}
 \centering
   $C_{2h}$&$ E$ &$ C_{2}$ &$\sigma_{h}$ & $i$ \\
             \hline
  $ \chi _{\rm{vector}}$&{ 3} &-1 & 1 & -3 \\
  $ \chi _{\rm{equivalent}}$ &{ 6} &0 & 6 & 0 \\
   $ \chi _{\rm{vibration}}$ &18 & 0 & 6 & 0\\
\end{tabular}
 \end{ruledtabular}
  \end{table}

We classify the lattice vibrational modes of T'-MoS$_2$ at $\Gamma$  by group theory according to the irreducible representations of C$_{2h}$. Characters of atomic displacement vector representations, primitive cell equivalent representations, and lattice vibration representations of T'-MoS$_2$ are shown in Table \ref{Reduced Representation}. These representations can be reduced into the irreducible representations summarized in Table \ref{character}:
\begin{align}
\label{eq4}
\Gamma _{\rm{vector}}&= 1{A_u} \oplus 2{B_u},\\
\label{eq5}
\Gamma _{\rm{equivalent}}& = 3{A_g}\oplus3{B_u},\\
\label{eq6}
\Gamma _{\rm{vibration}} &=\Gamma_{\rm{equivalent}}\otimes\Gamma_{\rm{vector}}\nonumber\\
                                  &=(3{A_g}\oplus3{B_u})\otimes( 1{A_u} \oplus 2{B_u})\nonumber \\
                                  &=3{A_u} \oplus 6{B_u} \oplus 3{B_g} \oplus 6{A_g},
\end{align}
where $\Gamma _{\rm{vector}}$, $\Gamma _{\rm{equivalent}}$, and $\Gamma _{\rm{vibration}}$ are the symmetry representations of atomic displacement vector, the equivalent representations of the primitive cell and  the symmetry representations of lattice vibration at the zone center of BZ, respectively. The symmetry representation of lattice vibration is equal to the direct product of  the symmetry representations of atomic displacement vector and the equivalent representations of the primitive cell.\cite{Dresselhaus2007group}

This symmetry representation of lattice vibration includes eighteen phonon modes entirely and can be further decomposed into the representations of acoustic and optical modes as follows:
 \begin{align}
   \label{eq7}
    \Gamma _{\rm{acoustic}}& =A_{u} \oplus 2B_{u},\\
    \label{eq8}
    \Gamma _{\rm{optical}} &= 2{A_u} \oplus 4{B_u} \oplus 6{A_g} \oplus 3{B_g},
  \end{align}
  where the acoustic modes include one $A_{u}$ and two $B_{u}$  modes,  all their frequencies are identical to zero; the rest of the fifteen nonzero frequency modes belong to optical modes. The six optical modes of odd parity (2$A_{u}$ and 4$B_{u}$) are IR active, the other nine optical modes of even parity (6$A_{g}$ and 3$B_{g}$) are R active. The R and IR modes are mutually exclusive in T'-MoS$_2$ phase because of the presence of inversion symmetry in the crystal. It is also worth pointing out that the above symmetry analyses is suitable for all T' phase of 2D TMD, namely, T'-MX$_2$ with M=(Mo, W) and X=(S, Se, and Te). For easy identifying T'-MoS$_2$ from a Raman optical spectral experiment, we compare the R modes of T'-MoS$_2$ with those of  H- and T-MoS$_2$.\cite{Cai2014, Zhang2015} It can be found that both H and T phase of MoS$_2$ have two-dimensional $E$ ($E'$ and $E_{g}$) modes, while T' phase has only one-dimensional modes, no two-dimensional $E$ mode. This means that if one detects the $E$ mode in a Raman optical spectral experiment on a MoS$_2$ monolayer, it coud not be in T' phase. In addition, since the presence of  inversion symmetry both in atomic structures of  T- and T'-MoS$_2$, according to exclusion principle, the R modes in these two phase must be $g$ modes, where T phase has both one- and two-dimensional $g$ modes ($A_1g$ and $E_{g}$) but T' phase has only one-dimensional $g$ modes ($A_{g}$ and $B_{g}$). In H-MoS$_2$, however, there is no inversion symmetry and thus no $g$ or $u$ mode. Thus, we may draw the conclusion that if one finds some one-dimensional but no two-dimensional R modes of $g$ symmetry in a Raman optical spectral experiment on a MoS$_2$ monolayer, then this  MoS$_2$ monolayer is probable in T' phase.
\subsubsection{Eigenfrequency and eigenvector of optical modes}
\begin{figure}[htbp]
\centering
\includegraphics[scale=0.26]{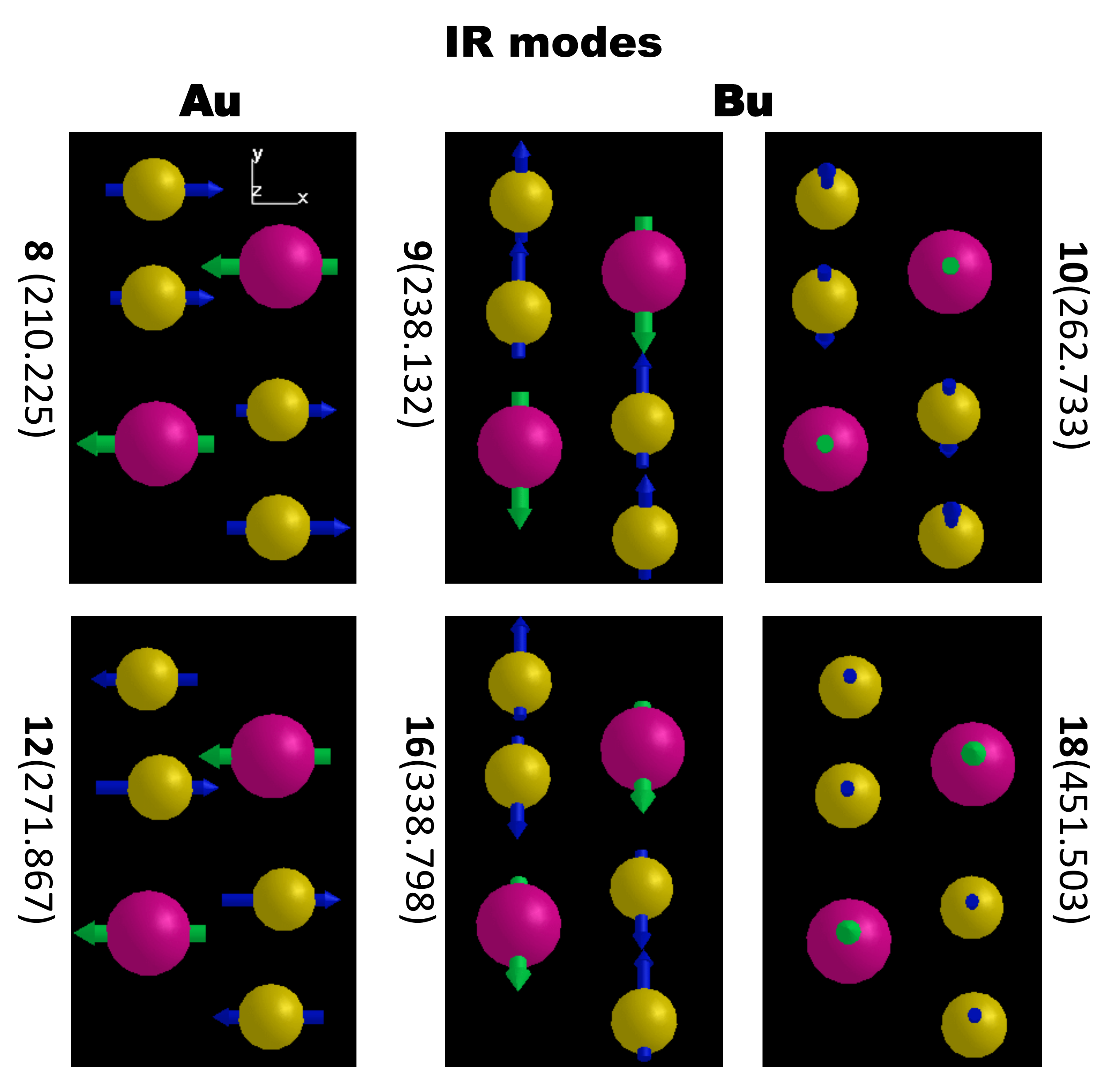}
\caption{\label{IR-modes-MoS2} (Color online) The eigenvectors of IR modes in T'-MoS$_2$.}
\end{figure}

\begin{figure}[htbp]
\centering
\includegraphics[scale=0.26]{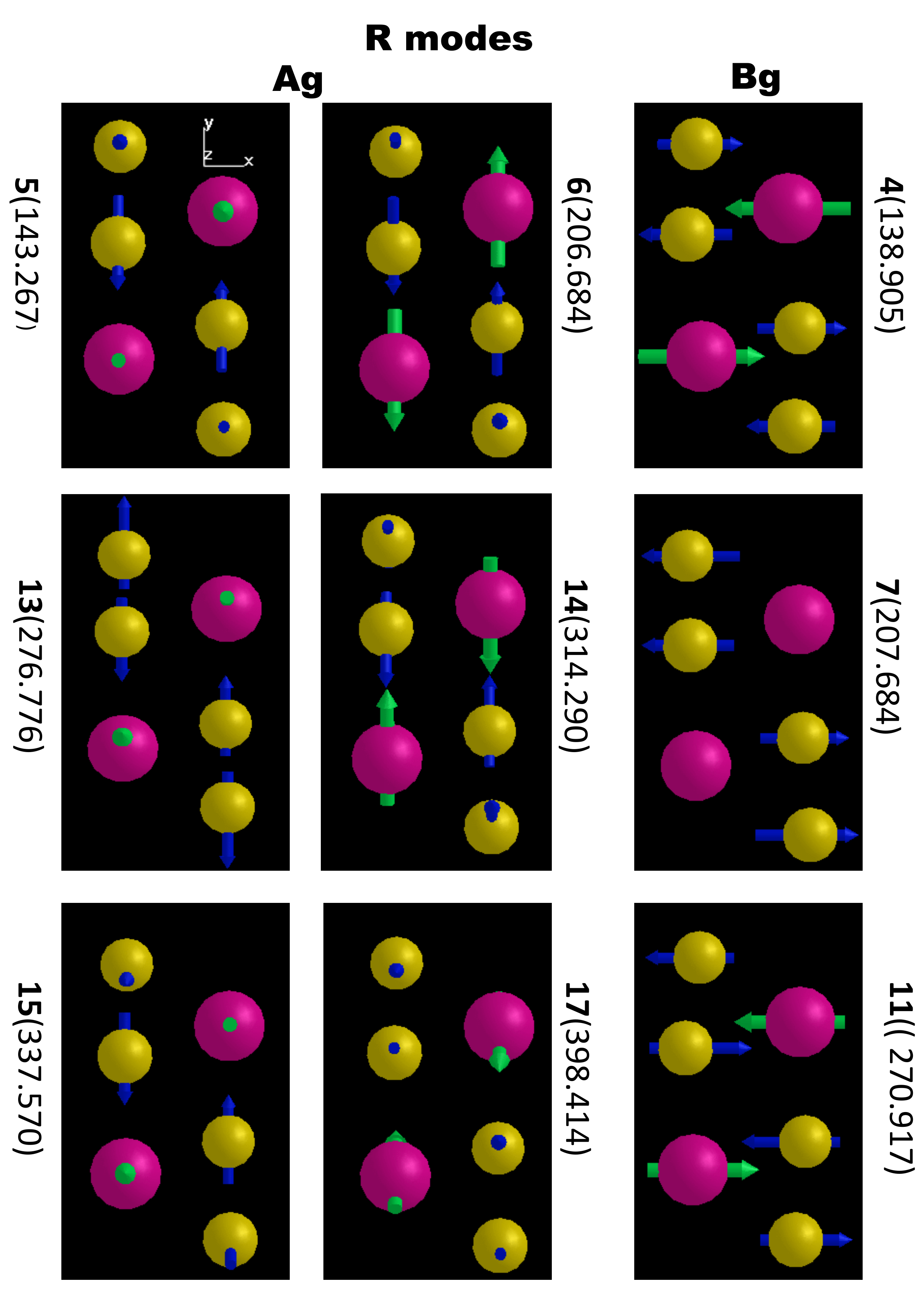}
\caption{\label{R-modes-MoS2} (Color online) The eigenvectors of R modes in T'-MoS$_2$.}
\end{figure}

For comparing quantitatively with optical spectra experiments, we compute the eigenfrequency of the fifteen optical modes by phonopy. In Table \ref{IR-Raman}, the fifteen optical modes with frequency are grouped by their irreducible representations and optical activity, where bold Arabic numbers represent the optical modes, which ordering are according to their frequencies from low to high. T'-MoS$_2$ structure can be identified and characterized based on Table \ref{IR-Raman} in future optical spectra experiments.
\begin{table}[htbp]
\centering
\begin{ruledtabular}
\caption{\label{IR-Raman}Classification of the fifteen optical modes in T'-MoS$_2$ according to irreducible representations and optical activity, where the optical modes are denoted by bold Arabic number from \textbf{4} to \textbf{18}, their frequencies are given in parentheses in the unit of cm$^{-1}$.}
 \begin{tabular}{cc|cc}
 \multicolumn{2}{c|}{Raman modes}&\multicolumn{2}{c}{Infrared modes} \\
 \hline
 \centering
 $A_{\rm{g}}$         &$ B _{\rm{g}}$      &$A_{\rm{u}}$         & $B _{\rm{u}}$\\
 \hline
 \textbf{ 5}(143.267)  &\textbf{ 4}(138.905)&\textbf{ 8}(210.225) &\textbf{ 9}(238.132)\\
 \textbf{ 6}(206.684)  &\textbf{ 7}(207.684)&\textbf{12}(271.867)&\textbf{10}(262.733)\\
 \textbf{13}(276.776)&\textbf{11}(270.917)&                               &\textbf{16}(338.798)\\
 \textbf{14}(314.290)&                               &                               &\textbf{18}(451.503)\\
 \textbf{15}(337.570)&                               &                                &\\
 \textbf{17}(398.414)&                               &                                &\\
 \end{tabular}\
 \end{ruledtabular}
 \end{table}
Besides, the vibrational eigenvector of the IR and R modes are also illustrated in Fig. \ref{IR-modes-MoS2} and Fig. \ref{R-modes-MoS2}. For IR modes, as can be seen in Fig. \ref{IR-modes-MoS2}, both the two $A_{u}$ modes \textbf{8}(210.225) and \textbf{12}(271.867) are vibrating along in-plane directions, while only one $B_{u}$ mode \textbf{18}(451.503) which is of the highest frequency, vibrates perpendicular to crystal plane. For IR modes, from Fig. \ref{R-modes-MoS2} we find that all the three $B_{g}$ modes \textbf{4}(138.905), \textbf{7}(207.684) and \textbf{11}(270.917) are in-plane vibrations, while none of the six $A_{g}$ modes is vibrating along purely in-plane or out-plane direction. The vibration direction of IR and R active modes is vital for setting the incident and detection directions as well as the polarization of the light used in optical spectra experiments.
\subsection{The Electronic Band Gap of  T'-MoS$_{2}$}
There has been a discrepancy about the band gap of T'-MoS$_2$  presented in recent literature.\cite{Kan2014, Qian2014, gao2015charge} Kan et al. performed spin-polarized DFT calculations with GGA-PBE and with HSE06 to investigate the band structure of  monolayer of T'-MoS$_2$. They pointed out that the structural distortions of  ZT-MoS$_2$ lead to the opening of a direct gap of 0.022 or 0.23 eV. The band gap obtained by DFT with GGA-PBE is significantly different from that obtained with HSE06, the latter is ten times greater than the former.\cite{Kan2014} Qian et al. later found that 1T'-MoS2 (i.e., T'-MoS2) was a semiconductor with a band gap of 0.1 eV based on many-body perturbation theory within the GW approximation.\cite{Qian2014} However, Gao et al.'s calculation by DFT based on Dmol3 software showed that T'-MoS$_2$  was a semiconductor with a very narrow band gap of 0.006eV.\cite{gao2015charge} Generally speaking, the band gap is underestimated by GGA-PBE but overestimated by HSE06.Thus, the band gap of 0.1 eV obtained by Qian et al.\cite{Qian2014} should be closer to the real value since this value is in between the two results: 0.022 eV and 0.23 eV, where the former is calculated by DFT with GGA-PBE and the latter is with HSE06. As to the band gap of 0.006 eV \cite{gao2015charge},  it may seriously underestimate the band gap since it is far lower than the underestimated result 0.022 eV.
\begin{figure}[htbp]
          \centering
          \includegraphics[scale=0.32]{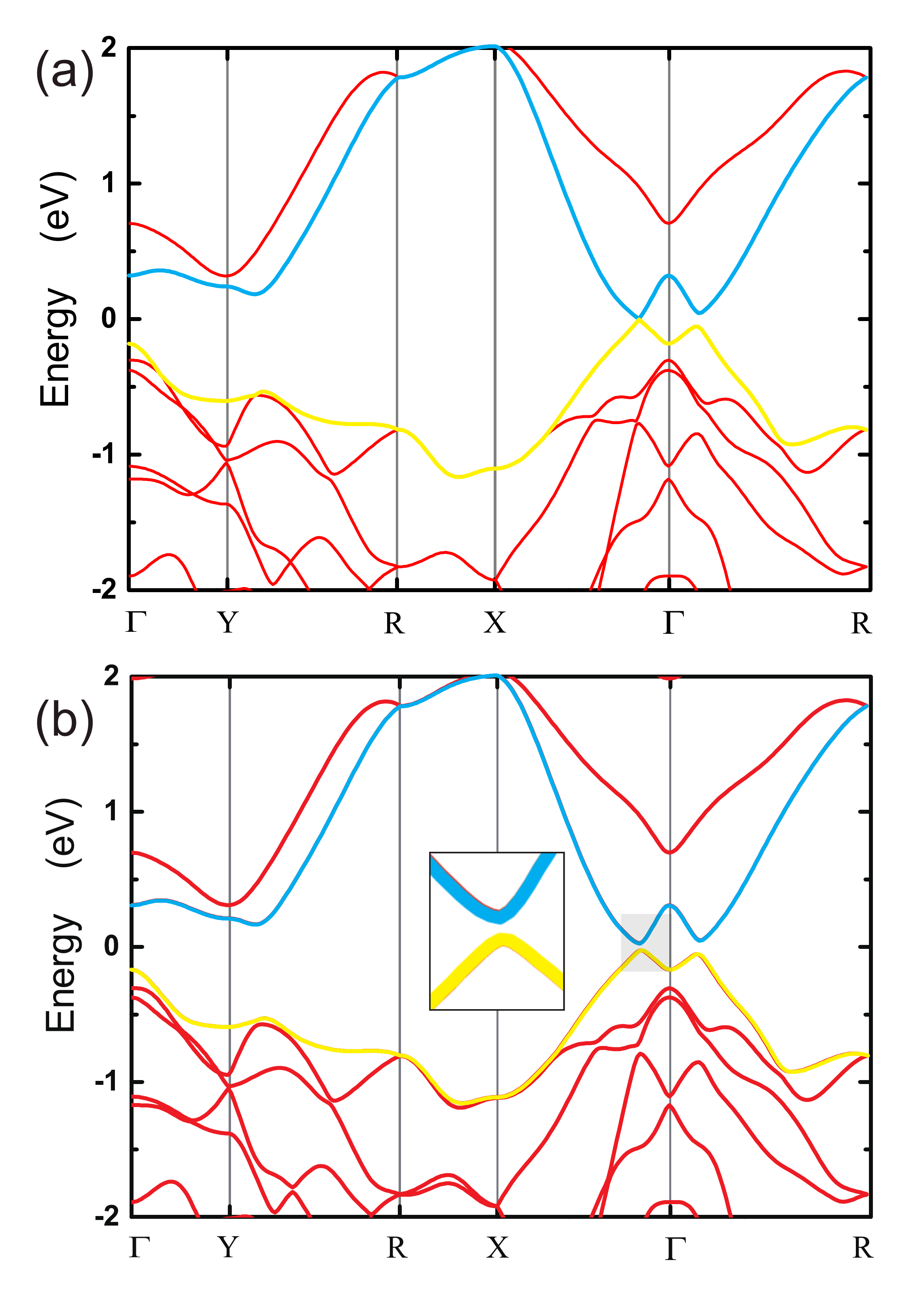}
           \caption{\label{electric-band-1}(Color online) Electronic band structure of T'-MoS$_2$  before (a) and after (b) considering spin-orbit coupling, and the vacuum level is set to zero. The band gap is opened by spin-orbit coupling.}
        \end{figure}
To examine our inference, we calculate the electronic band structure of T'-MoS$_2$ by DFT with GGA-PBE to explore the effect of spin-orbit coupling. Fig. \ref{electric-band-1} shows the electronic band structure of  T'-MoS$_2$ without (a) and with (b) the consideration of the spin-orbit coupling. Comparing  Fig. \ref{electric-band-1}(a) and \ref{electric-band-1}(b), one can find that the electronic band dispersion curves in the two cases are almost as the same in general, but the crucial difference in detail occurs near the Fermi line. In the former case, there seems no band gap, while in the latter case, the band gap does present, and is equal to 0.048 eV. This result implies that it is the spin-orbit coupling opens or widens out the band gap. Besides, we also find that the band structure without consider the spin-orbit agrees well with that of Gao et al.'s especially in the vicinity of the Fermi line.

\begin{figure}[htbp]
          \centering
          \includegraphics[scale=0.52]{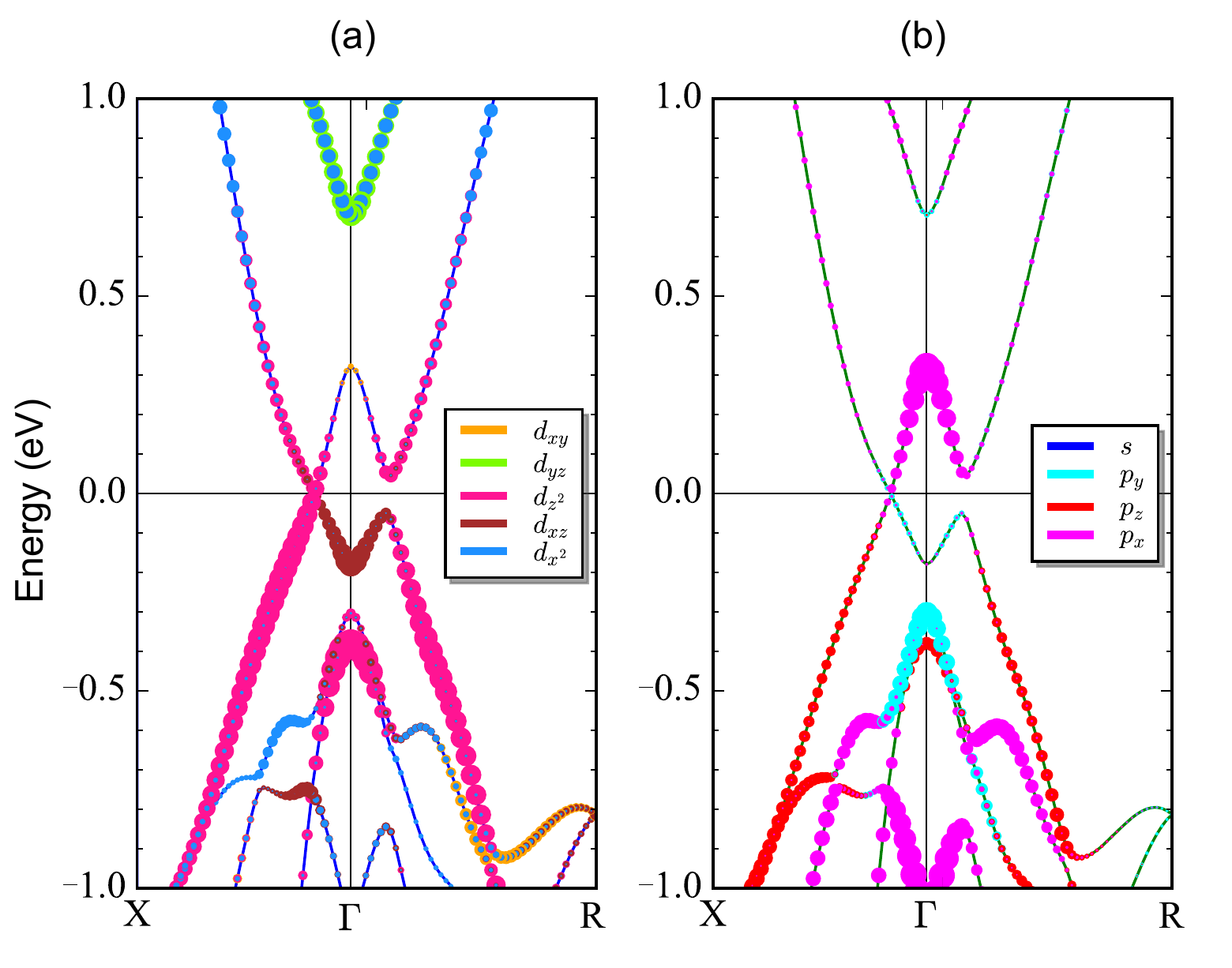}
           \caption{\label{pband}(Color online) Orbital-projected fine band structures of (a) Mo and (b) S atoms around the $\Gamma$ point without considering spin-orbit coupling. The diameter of circle indicates the weight of components.}
 \end{figure}

 To determine whether there exists a very small band gap without spin orbital interaction, we plot the orbital-projected band structures\cite{Wang2015Role} of Mo and S atom around the $\Gamma$ point without considering spin-orbit coupling. As can be seen in Fig. \ref{pband}, for both Mo and S atom, the two bands meeting at the Fermi lime cross each other directly without any avoiding.This direct crossing shows that there is exactly no band gap existing in electronic band structure of T'-MoS$_2$ if the spin-orbit interaction is neglected. Thus, we can conclude that the interaction which lifting the degeneracy of electron states at the Fermi line and opening the band gap is the spin-orbit coupling. In addition, we recalculate the electronic band structure of  T'-MoS$_2$ by DFT with HSE06 and the obtained band gap is about 0.153 eV, which falls in between those obtained in Refs [\onlinecite{Kan2014} and \onlinecite{Qian2014}]. Thus far, we come to conclusion that T'-MoS$_2$ must be a semiconductor of a narrow gap, while Gao et al.'s calculation may have not included the spin-orbit interaction.

\section{Conclusions}

In conclusion, we have performed first-principles investigation on the structure, physical stability, optical modes and electronic band gap of T'-MoS$_2$. Our simulated STM images of MoS$_2$ monolayer are in good agreement with previous experimental results. Moreover, we have found unexpectedly that the simulated STM images of T'-MoS$_2$ vary significantly with the scanning distance. This variation should be considered in the structural identification from experimental STM images. Furthermore, the dramatic similarity between the simulated STM images of T'-MoS$_2$ with that of earlier experimental study means that T'-MoS$_2$ may have been observed in experiment but was mistaken for the intercalation compound K$_{x}$(H$_{2}$O)$_{y}$MoS$_{2}$. If so, T'-MoS$_2$ should be stable in structure. To verify its physical stability, the thermal and mechanical stability of T'-MoS$_2$ have explored by AIMD simulations and elastic constants fitting and the results are affirmative. In addition, the lattice dynamic stability of T'-MoS$_2$ is also confirmed by the absence of imaginary frequency in our phonon dispersions relations. Therefore, the physical stability of T'-MoS$_2$ has been verified finally. Besides, we have made symmetry classification of optical modes and calculated their eigenfrequencies and eigenvectors, which provides an important guidance for further optical spectral study in experiments. Future work will investigate the intensity of R and IR spectra theoretically.

\begin{acknowledgments}
Y. C.  Liu is thankful to H. B. Niu for his help on first principles calculations. V.Wang. acknowledges the financial support of The Special Scientific Research Program of the Education Bureau of Shaanxi Province, China (Grant No. 15JK1531).
\end{acknowledgments}
\nocite{*}
\bibliographystyle{aipnum4-1}
\bibliography{References}

\begin{thebibliography}{62}%
\makeatletter
\providecommand \@ifxundefined [1]{%
 \@ifx{#1\undefined}
}%
\providecommand \@ifnum [1]{%
 \ifnum #1\expandafter \@firstoftwo
 \else \expandafter \@secondoftwo
 \fi
}%
\providecommand \@ifx [1]{%
 \ifx #1\expandafter \@firstoftwo
 \else \expandafter \@secondoftwo
 \fi
}%
\providecommand \natexlab [1]{#1}%
\providecommand \enquote  [1]{``#1''}%
\providecommand \bibnamefont  [1]{#1}%
\providecommand \bibfnamefont [1]{#1}%
\providecommand \citenamefont [1]{#1}%
\providecommand \href@noop [0]{\@secondoftwo}%
\providecommand \href [0]{\begingroup \@sanitize@url \@href}%
\providecommand \@href[1]{\@@startlink{#1}\@@href}%
\providecommand \@@href[1]{\endgroup#1\@@endlink}%
\providecommand \@sanitize@url [0]{\catcode `\\12\catcode `\$12\catcode
  `\&12\catcode `\#12\catcode `\^12\catcode `\_12\catcode `\%12\relax}%
\providecommand \@@startlink[1]{}%
\providecommand \@@endlink[0]{}%
\providecommand \url  [0]{\begingroup\@sanitize@url \@url }%
\providecommand \@url [1]{\endgroup\@href {#1}{\urlprefix }}%
\providecommand \urlprefix  [0]{URL }%
\providecommand \Eprint [0]{\href }%
\providecommand \doibase [0]{http://dx.doi.org/}%
\providecommand \selectlanguage [0]{\@gobble}%
\providecommand \bibinfo  [0]{\@secondoftwo}%
\providecommand \bibfield  [0]{\@secondoftwo}%
\providecommand \translation [1]{[#1]}%
\providecommand \BibitemOpen [0]{}%
\providecommand \bibitemStop [0]{}%
\providecommand \bibitemNoStop [0]{.\EOS\space}%
\providecommand \EOS [0]{\spacefactor3000\relax}%
\providecommand \BibitemShut  [1]{\csname bibitem#1\endcsname}%
\let\auto@bib@innerbib\@empty
\bibitem [{\citenamefont {Novoselov}\ \emph {et~al.}(2005)\citenamefont
  {Novoselov}, \citenamefont {Geim}, \citenamefont {Morozov}, \citenamefont
  {Jiang}, \citenamefont {Katsnelson}, \citenamefont {Grigorieva},
  \citenamefont {Dubonos},\ and\ \citenamefont {Firsov}}]{Novoselov2005}%
  \BibitemOpen
  \bibfield  {author} {\bibinfo {author} {\bibfnamefont {K.}~\bibnamefont
  {Novoselov}}, \bibinfo {author} {\bibfnamefont {A.~K.}\ \bibnamefont {Geim}},
  \bibinfo {author} {\bibfnamefont {S.}~\bibnamefont {Morozov}}, \bibinfo
  {author} {\bibfnamefont {D.}~\bibnamefont {Jiang}}, \bibinfo {author}
  {\bibfnamefont {M.}~\bibnamefont {Katsnelson}}, \bibinfo {author}
  {\bibfnamefont {I.}~\bibnamefont {Grigorieva}}, \bibinfo {author}
  {\bibfnamefont {S.}~\bibnamefont {Dubonos}}, \ and\ \bibinfo {author}
  {\bibfnamefont {A.}~\bibnamefont {Firsov}},\ }\href
  {http://www.nature.com/nature/journal/v438/n7065/abs/nature04233.html}
  {\bibfield  {journal} {\bibinfo  {journal} {Nature}\ }\textbf {\bibinfo
  {volume} {438}},\ \bibinfo {pages} {197} (\bibinfo {year}
  {2005})}\BibitemShut {NoStop}%
\bibitem [{\citenamefont {Zhang}\ \emph {et~al.}(2005)\citenamefont {Zhang},
  \citenamefont {Tan}, \citenamefont {Stormer},\ and\ \citenamefont
  {Kim}}]{Zhang2005}%
  \BibitemOpen
  \bibfield  {author} {\bibinfo {author} {\bibfnamefont {Y.}~\bibnamefont
  {Zhang}}, \bibinfo {author} {\bibfnamefont {Y.-W.}\ \bibnamefont {Tan}},
  \bibinfo {author} {\bibfnamefont {H.~L.}\ \bibnamefont {Stormer}}, \ and\
  \bibinfo {author} {\bibfnamefont {P.}~\bibnamefont {Kim}},\ }\href
  {http://www.nature.com/nature/journal/v438/n7065/abs/nature04235.html}
  {\bibfield  {journal} {\bibinfo  {journal} {Nature}\ }\textbf {\bibinfo
  {volume} {438}},\ \bibinfo {pages} {201} (\bibinfo {year}
  {2005})}\BibitemShut {NoStop}%
\bibitem [{\citenamefont {Liao}\ \emph {et~al.}(2010)\citenamefont {Liao},
  \citenamefont {Lin}, \citenamefont {Bao}, \citenamefont {Cheng},
  \citenamefont {Bai}, \citenamefont {Liu}, \citenamefont {Qu}, \citenamefont
  {Wang}, \citenamefont {Huang},\ and\ \citenamefont {Duan}}]{Liao2010}%
  \BibitemOpen
  \bibfield  {author} {\bibinfo {author} {\bibfnamefont {L.}~\bibnamefont
  {Liao}}, \bibinfo {author} {\bibfnamefont {Y.-C.}\ \bibnamefont {Lin}},
  \bibinfo {author} {\bibfnamefont {M.}~\bibnamefont {Bao}}, \bibinfo {author}
  {\bibfnamefont {R.}~\bibnamefont {Cheng}}, \bibinfo {author} {\bibfnamefont
  {J.}~\bibnamefont {Bai}}, \bibinfo {author} {\bibfnamefont {Y.}~\bibnamefont
  {Liu}}, \bibinfo {author} {\bibfnamefont {Y.}~\bibnamefont {Qu}}, \bibinfo
  {author} {\bibfnamefont {K.~L.}\ \bibnamefont {Wang}}, \bibinfo {author}
  {\bibfnamefont {Y.}~\bibnamefont {Huang}}, \ and\ \bibinfo {author}
  {\bibfnamefont {X.}~\bibnamefont {Duan}},\ }\href
  {http://www.nature.com/nature/journal/v467/n7313/full/nature09405.html}
  {\bibfield  {journal} {\bibinfo  {journal} {Nature}\ }\textbf {\bibinfo
  {volume} {467}},\ \bibinfo {pages} {305} (\bibinfo {year}
  {2010})}\BibitemShut {NoStop}%
\bibitem [{\citenamefont {Schwierz}(2010)}]{Schwierz2010}%
  \BibitemOpen
  \bibfield  {author} {\bibinfo {author} {\bibfnamefont {F.}~\bibnamefont
  {Schwierz}},\ }\href
  {http://www.nature.com/nnano/journal/v5/n7/full/nnano.2010.89.html}
  {\bibfield  {journal} {\bibinfo  {journal} {Nat. Nanotechnol.}\ }\textbf
  {\bibinfo {volume} {5}},\ \bibinfo {pages} {487} (\bibinfo {year}
  {2010})}\BibitemShut {NoStop}%
\bibitem [{\citenamefont {Mak}\ \emph {et~al.}(2010)\citenamefont {Mak},
  \citenamefont {Lee}, \citenamefont {Hone}, \citenamefont {Shan},\ and\
  \citenamefont {Heinz}}]{Mak2010}%
  \BibitemOpen
  \bibfield  {author} {\bibinfo {author} {\bibfnamefont {K.~F.}\ \bibnamefont
  {Mak}}, \bibinfo {author} {\bibfnamefont {C.}~\bibnamefont {Lee}}, \bibinfo
  {author} {\bibfnamefont {J.}~\bibnamefont {Hone}}, \bibinfo {author}
  {\bibfnamefont {J.}~\bibnamefont {Shan}}, \ and\ \bibinfo {author}
  {\bibfnamefont {T.~F.}\ \bibnamefont {Heinz}},\ }\href
  {http://journals.aps.org/prl/abstract/10.1103/PhysRevLett.105.136805}
  {\bibfield  {journal} {\bibinfo  {journal} {Phys. Rev. Lett.}\ }\textbf
  {\bibinfo {volume} {105}},\ \bibinfo {pages} {136805} (\bibinfo {year}
  {2010})}\BibitemShut {NoStop}%
\bibitem [{\citenamefont {Wu}\ \emph {et~al.}(2011)\citenamefont {Wu},
  \citenamefont {Lin}, \citenamefont {Bol}, \citenamefont {Jenkins},
  \citenamefont {Xia}, \citenamefont {Farmer}, \citenamefont {Zhu},\ and\
  \citenamefont {Avouris}}]{Wu2011}%
  \BibitemOpen
  \bibfield  {author} {\bibinfo {author} {\bibfnamefont {Y.}~\bibnamefont
  {Wu}}, \bibinfo {author} {\bibfnamefont {Y.-m.}\ \bibnamefont {Lin}},
  \bibinfo {author} {\bibfnamefont {A.~A.}\ \bibnamefont {Bol}}, \bibinfo
  {author} {\bibfnamefont {K.~A.}\ \bibnamefont {Jenkins}}, \bibinfo {author}
  {\bibfnamefont {F.}~\bibnamefont {Xia}}, \bibinfo {author} {\bibfnamefont
  {D.~B.}\ \bibnamefont {Farmer}}, \bibinfo {author} {\bibfnamefont
  {Y.}~\bibnamefont {Zhu}}, \ and\ \bibinfo {author} {\bibfnamefont
  {P.}~\bibnamefont {Avouris}},\ }\href
  {http://www.nature.com/nature/journal/v472/n7341} {\bibfield  {journal}
  {\bibinfo  {journal} {Nature}\ }\textbf {\bibinfo {volume} {472}},\ \bibinfo
  {pages} {74} (\bibinfo {year} {2011})}\BibitemShut {NoStop}%
\bibitem [{\citenamefont {Kan}\ \emph {et~al.}(2014)\citenamefont {Kan},
  \citenamefont {Wang}, \citenamefont {Li}, \citenamefont {Zhang},
  \citenamefont {Li}, \citenamefont {Kawazoe}, \citenamefont {Sun},\ and\
  \citenamefont {Jena}}]{Kan2014}%
  \BibitemOpen
  \bibfield  {author} {\bibinfo {author} {\bibfnamefont {M.}~\bibnamefont
  {Kan}}, \bibinfo {author} {\bibfnamefont {J.~Y.}\ \bibnamefont {Wang}},
  \bibinfo {author} {\bibfnamefont {X.~W.}\ \bibnamefont {Li}}, \bibinfo
  {author} {\bibfnamefont {S.~H.}\ \bibnamefont {Zhang}}, \bibinfo {author}
  {\bibfnamefont {Y.~W.}\ \bibnamefont {Li}}, \bibinfo {author} {\bibfnamefont
  {Y.}~\bibnamefont {Kawazoe}}, \bibinfo {author} {\bibfnamefont
  {Q.}~\bibnamefont {Sun}}, \ and\ \bibinfo {author} {\bibfnamefont
  {P.}~\bibnamefont {Jena}},\ }\href {http://dx.doi.org/10.1021/jp4076355}
  {\bibfield  {journal} {\bibinfo  {journal} {J. Phys. Chem. C}\ }\textbf
  {\bibinfo {volume} {118}},\ \bibinfo {pages} {1515} (\bibinfo {year}
  {2014})}\BibitemShut {NoStop}%
\bibitem [{\citenamefont {Li}\ \emph {et~al.}(2014)\citenamefont {Li},
  \citenamefont {Guo}, \citenamefont {Zhang},\ and\ \citenamefont
  {Zhang}}]{Li2014Gapless}%
  \BibitemOpen
  \bibfield  {author} {\bibinfo {author} {\bibfnamefont {W.}~\bibnamefont
  {Li}}, \bibinfo {author} {\bibfnamefont {M.}~\bibnamefont {Guo}}, \bibinfo
  {author} {\bibfnamefont {G.}~\bibnamefont {Zhang}}, \ and\ \bibinfo {author}
  {\bibfnamefont {Y.-W.}\ \bibnamefont {Zhang}},\ }\href {\doibase
  10.1103/PhysRevB.89.205402} {\bibfield  {journal} {\bibinfo  {journal} {Phys.
  Rev. B}\ }\textbf {\bibinfo {volume} {89}},\ \bibinfo {pages} {205402}
  (\bibinfo {year} {2014})}\BibitemShut {NoStop}%
\bibitem [{\citenamefont {Wang}\ \emph {et~al.}(2012)\citenamefont {Wang},
  \citenamefont {Kalantar-Zadeh}, \citenamefont {Kis}, \citenamefont
  {Coleman},\ and\ \citenamefont {Strano}}]{Wang2012}%
  \BibitemOpen
  \bibfield  {author} {\bibinfo {author} {\bibfnamefont {Q.~H.}\ \bibnamefont
  {Wang}}, \bibinfo {author} {\bibfnamefont {K.}~\bibnamefont
  {Kalantar-Zadeh}}, \bibinfo {author} {\bibfnamefont {A.}~\bibnamefont {Kis}},
  \bibinfo {author} {\bibfnamefont {J.~N.}\ \bibnamefont {Coleman}}, \ and\
  \bibinfo {author} {\bibfnamefont {M.~S.}\ \bibnamefont {Strano}},\ }\href
  {http://www.nature.com/nnano/journal/v7/n11/full/nnano.2012.193.html}
  {\bibfield  {journal} {\bibinfo  {journal} {Nat. Nanotechnol.}\ }\textbf
  {\bibinfo {volume} {7}},\ \bibinfo {pages} {699} (\bibinfo {year}
  {2012})}\BibitemShut {NoStop}%
\bibitem [{\citenamefont {Py}\ and\ \citenamefont
  {Haering}(1983)}]{py1983structural}%
  \BibitemOpen
  \bibfield  {author} {\bibinfo {author} {\bibfnamefont {M.}~\bibnamefont
  {Py}}\ and\ \bibinfo {author} {\bibfnamefont {R.}~\bibnamefont {Haering}},\
  }\href {http://www.nrcresearchpress.com/doi/abs/10.1139/p83-013#.VuI6YPmPMRE}
  {\bibfield  {journal} {\bibinfo  {journal} {Can. J. Phys.}\ }\textbf
  {\bibinfo {volume} {61}},\ \bibinfo {pages} {76} (\bibinfo {year}
  {1983})}\BibitemShut {NoStop}%
\bibitem [{\citenamefont {F.~Wypych}\ and\ \citenamefont
  {Prins}(1998)}]{Wypych1998}%
  \BibitemOpen
  \bibfield  {author} {\bibinfo {author} {\bibfnamefont {T.~W.}\ \bibnamefont
  {F.~Wypych}}\ and\ \bibinfo {author} {\bibfnamefont {R.}~\bibnamefont
  {Prins}},\ }\href {\doibase 10.1021/cm970402e} {\bibfield  {journal}
  {\bibinfo  {journal} {Chem. Mater.}\ }\textbf {\bibinfo {volume} {10}},\
  \bibinfo {pages} {723} (\bibinfo {year} {1998})}\BibitemShut {NoStop}%
\bibitem [{\citenamefont {Ataca}, \citenamefont {Sahin},\ and\ \citenamefont
  {Ciraci}(2012)}]{ataca2012stable}%
  \BibitemOpen
  \bibfield  {author} {\bibinfo {author} {\bibfnamefont {C.}~\bibnamefont
  {Ataca}}, \bibinfo {author} {\bibfnamefont {H.}~\bibnamefont {Sahin}}, \ and\
  \bibinfo {author} {\bibfnamefont {S.}~\bibnamefont {Ciraci}},\ }\href
  {http://pubs.acs.org/doi/abs/10.1021/jp212558p} {\bibfield  {journal}
  {\bibinfo  {journal} {J. Phys. Chem. C}\ }\textbf {\bibinfo {volume} {116}},\
  \bibinfo {pages} {8983} (\bibinfo {year} {2012})}\BibitemShut {NoStop}%
\bibitem [{\citenamefont {Benavente}\ \emph {et~al.}(2002)\citenamefont
  {Benavente}, \citenamefont {Santa~Ana}, \citenamefont {Mendiz{\'a}bal},\ and\
  \citenamefont {Gonz{\'a}lez}}]{benavente2002intercalation}%
  \BibitemOpen
  \bibfield  {author} {\bibinfo {author} {\bibfnamefont {E.}~\bibnamefont
  {Benavente}}, \bibinfo {author} {\bibfnamefont {M.}~\bibnamefont
  {Santa~Ana}}, \bibinfo {author} {\bibfnamefont {F.}~\bibnamefont
  {Mendiz{\'a}bal}}, \ and\ \bibinfo {author} {\bibfnamefont {G.}~\bibnamefont
  {Gonz{\'a}lez}},\ }\href
  {http://www.sciencedirect.com/science/article/pii/S0010854501003927}
  {\bibfield  {journal} {\bibinfo  {journal} {Coord. Chem. Rev.}\ }\textbf
  {\bibinfo {volume} {224}},\ \bibinfo {pages} {87} (\bibinfo {year}
  {2002})}\BibitemShut {NoStop}%
\bibitem [{\citenamefont {Lee}\ \emph {et~al.}(2012)\citenamefont {Lee},
  \citenamefont {Zhang}, \citenamefont {Zhang}, \citenamefont {Chang},
  \citenamefont {Lin}, \citenamefont {Chang}, \citenamefont {Yu}, \citenamefont
  {Wang}, \citenamefont {Chang}, \citenamefont {Li},\ and\ \citenamefont
  {Lin}}]{Lee2012}%
  \BibitemOpen
  \bibfield  {author} {\bibinfo {author} {\bibfnamefont {Y.-H.}\ \bibnamefont
  {Lee}}, \bibinfo {author} {\bibfnamefont {X.-Q.}\ \bibnamefont {Zhang}},
  \bibinfo {author} {\bibfnamefont {W.}~\bibnamefont {Zhang}}, \bibinfo
  {author} {\bibfnamefont {M.-T.}\ \bibnamefont {Chang}}, \bibinfo {author}
  {\bibfnamefont {C.-T.}\ \bibnamefont {Lin}}, \bibinfo {author} {\bibfnamefont
  {K.-D.}\ \bibnamefont {Chang}}, \bibinfo {author} {\bibfnamefont {Y.-C.}\
  \bibnamefont {Yu}}, \bibinfo {author} {\bibfnamefont {J.~T.-W.}\ \bibnamefont
  {Wang}}, \bibinfo {author} {\bibfnamefont {C.-S.}\ \bibnamefont {Chang}},
  \bibinfo {author} {\bibfnamefont {L.-J.}\ \bibnamefont {Li}}, \ and\ \bibinfo
  {author} {\bibfnamefont {T.-W.}\ \bibnamefont {Lin}},\ }\href {\doibase
  10.1002/adma.201104798} {\bibfield  {journal} {\bibinfo  {journal} {Adv.
  Mater.}\ }\textbf {\bibinfo {volume} {24}},\ \bibinfo {pages} {2320}
  (\bibinfo {year} {2012})}\BibitemShut {NoStop}%
\bibitem [{\citenamefont {Qin}\ \emph {et~al.}(1991)\citenamefont {Qin},
  \citenamefont {Yang}, \citenamefont {Frindt},\ and\ \citenamefont
  {Irwin}}]{Qin1991}%
  \BibitemOpen
  \bibfield  {author} {\bibinfo {author} {\bibfnamefont {X.~R.}\ \bibnamefont
  {Qin}}, \bibinfo {author} {\bibfnamefont {D.}~\bibnamefont {Yang}}, \bibinfo
  {author} {\bibfnamefont {R.~F.}\ \bibnamefont {Frindt}}, \ and\ \bibinfo
  {author} {\bibfnamefont {J.~C.}\ \bibnamefont {Irwin}},\ }\href {\doibase
  10.1103/PhysRevB.44.3490} {\bibfield  {journal} {\bibinfo  {journal} {Phys.
  Rev. B}\ }\textbf {\bibinfo {volume} {44}},\ \bibinfo {pages} {3490}
  (\bibinfo {year} {1991})}\BibitemShut {NoStop}%
\bibitem [{\citenamefont {Radisavljevic}\ \emph {et~al.}(2011)\citenamefont
  {Radisavljevic}, \citenamefont {Radenovic}, \citenamefont {Brivio},
  \citenamefont {Giacometti},\ and\ \citenamefont {Kis}}]{Radisavljevic2011}%
  \BibitemOpen
  \bibfield  {author} {\bibinfo {author} {\bibfnamefont {B.}~\bibnamefont
  {Radisavljevic}}, \bibinfo {author} {\bibfnamefont {A.}~\bibnamefont
  {Radenovic}}, \bibinfo {author} {\bibfnamefont {J.}~\bibnamefont {Brivio}},
  \bibinfo {author} {\bibfnamefont {V.}~\bibnamefont {Giacometti}}, \ and\
  \bibinfo {author} {\bibfnamefont {A.}~\bibnamefont {Kis}},\ }\href
  {http://www.nature.com/nnano/journal/v6/n3/full/nnano.2010.279.html}
  {\bibfield  {journal} {\bibinfo  {journal} {Nat. Nanotechnol.}\ }\textbf
  {\bibinfo {volume} {6}},\ \bibinfo {pages} {147} (\bibinfo {year}
  {2011})}\BibitemShut {NoStop}%
\bibitem [{\citenamefont {Kappera}\ \emph {et~al.}(2014)\citenamefont
  {Kappera}, \citenamefont {Voiry}, \citenamefont {Yalcin}, \citenamefont
  {Jen}, \citenamefont {Acerce}, \citenamefont {Torrel}, \citenamefont
  {Branch}, \citenamefont {Lei}, \citenamefont {Chen}, \citenamefont {Najmaei},
  \citenamefont {Lou}, \citenamefont {Ajayan}, \citenamefont {Gupta},
  \citenamefont {Mohite},\ and\ \citenamefont {Chhowalla}}]{Kappera2014}%
  \BibitemOpen
  \bibfield  {author} {\bibinfo {author} {\bibfnamefont {R.}~\bibnamefont
  {Kappera}}, \bibinfo {author} {\bibfnamefont {D.}~\bibnamefont {Voiry}},
  \bibinfo {author} {\bibfnamefont {S.~E.}\ \bibnamefont {Yalcin}}, \bibinfo
  {author} {\bibfnamefont {W.}~\bibnamefont {Jen}}, \bibinfo {author}
  {\bibfnamefont {M.}~\bibnamefont {Acerce}}, \bibinfo {author} {\bibfnamefont
  {S.}~\bibnamefont {Torrel}}, \bibinfo {author} {\bibfnamefont
  {B.}~\bibnamefont {Branch}}, \bibinfo {author} {\bibfnamefont
  {S.}~\bibnamefont {Lei}}, \bibinfo {author} {\bibfnamefont {W.}~\bibnamefont
  {Chen}}, \bibinfo {author} {\bibfnamefont {S.}~\bibnamefont {Najmaei}},
  \bibinfo {author} {\bibfnamefont {J.}~\bibnamefont {Lou}}, \bibinfo {author}
  {\bibfnamefont {P.~M.}\ \bibnamefont {Ajayan}}, \bibinfo {author}
  {\bibfnamefont {G.}~\bibnamefont {Gupta}}, \bibinfo {author} {\bibfnamefont
  {A.~D.}\ \bibnamefont {Mohite}}, \ and\ \bibinfo {author} {\bibfnamefont
  {M.}~\bibnamefont {Chhowalla}},\ }\href
  {http://scitation.aip.org/content/aip/journal/aplmater/2/9/10.1063/1.4896077}
  {\bibfield  {journal} {\bibinfo  {journal} {APL Mater.}\ }\textbf {\bibinfo
  {volume} {2}},\ \bibinfo {pages} {092516} (\bibinfo {year}
  {2014})}\BibitemShut {NoStop}%
\bibitem [{\citenamefont {Lin}\ \emph {et~al.}(2013)\citenamefont {Lin},
  \citenamefont {Dumcenco}, \citenamefont {Huang},\ and\ \citenamefont
  {Suenaga}}]{lin2013atomic}%
  \BibitemOpen
  \bibfield  {author} {\bibinfo {author} {\bibfnamefont {Y.-C.}\ \bibnamefont
  {Lin}}, \bibinfo {author} {\bibfnamefont {D.~O.}\ \bibnamefont {Dumcenco}},
  \bibinfo {author} {\bibfnamefont {Y.-S.}\ \bibnamefont {Huang}}, \ and\
  \bibinfo {author} {\bibfnamefont {K.}~\bibnamefont {Suenaga}},\ }\href
  {http://arxiv.org/abs/1310.2363} {\bibfield  {journal} {\bibinfo  {journal}
  {arXiv:1310.2363}\ } (\bibinfo {year} {2013})}\BibitemShut {NoStop}%
\bibitem [{\citenamefont {Shirodkar}\ and\ \citenamefont
  {Waghmare}(2014)}]{Shirodkar2014emergence}%
  \BibitemOpen
  \bibfield  {author} {\bibinfo {author} {\bibfnamefont {S.~N.}\ \bibnamefont
  {Shirodkar}}\ and\ \bibinfo {author} {\bibfnamefont {U.~V.}\ \bibnamefont
  {Waghmare}},\ }\href
  {http://journals.aps.org/prl/abstract/10.1103/PhysRevLett.112.157601}
  {\bibfield  {journal} {\bibinfo  {journal} {Phys. Rev. Lett.}\ }\textbf
  {\bibinfo {volume} {112}},\ \bibinfo {pages} {157601} (\bibinfo {year}
  {2014})}\BibitemShut {NoStop}%
\bibitem [{\citenamefont {Singh}, \citenamefont {Shirodkar},\ and\
  \citenamefont {Waghmare}(2015)}]{Singh2015}%
  \BibitemOpen
  \bibfield  {author} {\bibinfo {author} {\bibfnamefont {A.}~\bibnamefont
  {Singh}}, \bibinfo {author} {\bibfnamefont {S.~N.}\ \bibnamefont
  {Shirodkar}}, \ and\ \bibinfo {author} {\bibfnamefont {U.~V.}\ \bibnamefont
  {Waghmare}},\ }\href {http://stacks.iop.org/2053-1583/2/i=3/a=035013}
  {\bibfield  {journal} {\bibinfo  {journal} {2D Mater.}\ }\textbf {\bibinfo
  {volume} {2}},\ \bibinfo {pages} {035013} (\bibinfo {year}
  {2015})}\BibitemShut {NoStop}%
\bibitem [{\citenamefont {Qin}\ \emph {et~al.}(1992)\citenamefont {Qin},
  \citenamefont {Yang}, \citenamefont {Frindt},\ and\ \citenamefont
  {Irwin}}]{Qin1992}%
  \BibitemOpen
  \bibfield  {author} {\bibinfo {author} {\bibfnamefont {X.}~\bibnamefont
  {Qin}}, \bibinfo {author} {\bibfnamefont {D.}~\bibnamefont {Yang}}, \bibinfo
  {author} {\bibfnamefont {R.}~\bibnamefont {Frindt}}, \ and\ \bibinfo {author}
  {\bibfnamefont {J.}~\bibnamefont {Irwin}},\ }\href
  {http://www.sciencedirect.com/science/article/pii/030439919290334G}
  {\bibfield  {journal} {\bibinfo  {journal} {Ultramicroscopy}\ }\textbf
  {\bibinfo {volume} {42}},\ \bibinfo {pages} {630 } (\bibinfo {year}
  {1992})}\BibitemShut {NoStop}%
\bibitem [{\citenamefont {Heising}\ and\ \citenamefont
  {Kanatzidis}(1999)}]{heising1999structure}%
  \BibitemOpen
  \bibfield  {author} {\bibinfo {author} {\bibfnamefont {J.}~\bibnamefont
  {Heising}}\ and\ \bibinfo {author} {\bibfnamefont {M.~G.}\ \bibnamefont
  {Kanatzidis}},\ }\href {http://pubs.acs.org/doi/abs/10.1021/ja983043c}
  {\bibfield  {journal} {\bibinfo  {journal} {J. Am. Chem. Soc.}\ }\textbf
  {\bibinfo {volume} {121}},\ \bibinfo {pages} {638} (\bibinfo {year}
  {1999})}\BibitemShut {NoStop}%
\bibitem [{\citenamefont {Whangbo}\ and\ \citenamefont
  {Canadell}(1992)}]{Whangbo1992}%
  \BibitemOpen
  \bibfield  {author} {\bibinfo {author} {\bibfnamefont {M.~H.}\ \bibnamefont
  {Whangbo}}\ and\ \bibinfo {author} {\bibfnamefont {E.}~\bibnamefont
  {Canadell}},\ }\href {\doibase 10.1021/ja00050a044} {\bibfield  {journal}
  {\bibinfo  {journal} {J. Am. Chem. Soc.}\ }\textbf {\bibinfo {volume}
  {114}},\ \bibinfo {pages} {9587} (\bibinfo {year} {1992})}\BibitemShut
  {NoStop}%
\bibitem [{\citenamefont {Qian}\ \emph {et~al.}(2014)\citenamefont {Qian},
  \citenamefont {Liu}, \citenamefont {Fu},\ and\ \citenamefont
  {Li}}]{Qian2014}%
  \BibitemOpen
  \bibfield  {author} {\bibinfo {author} {\bibfnamefont {X.}~\bibnamefont
  {Qian}}, \bibinfo {author} {\bibfnamefont {J.}~\bibnamefont {Liu}}, \bibinfo
  {author} {\bibfnamefont {L.}~\bibnamefont {Fu}}, \ and\ \bibinfo {author}
  {\bibfnamefont {J.}~\bibnamefont {Li}},\ }\href {\doibase
  10.1126/science.1256815} {\bibfield  {journal} {\bibinfo  {journal}
  {Science}\ }\textbf {\bibinfo {volume} {346}},\ \bibinfo {pages} {1344}
  (\bibinfo {year} {2014})}\BibitemShut {NoStop}%
\bibitem [{\citenamefont {Eda}\ \emph {et~al.}(2012)\citenamefont {Eda},
  \citenamefont {Fujita}, \citenamefont {Yamaguchi}, \citenamefont {Voiry},
  \citenamefont {Chen},\ and\ \citenamefont {Chhowalla}}]{Eda2012}%
  \BibitemOpen
  \bibfield  {author} {\bibinfo {author} {\bibfnamefont {G.}~\bibnamefont
  {Eda}}, \bibinfo {author} {\bibfnamefont {T.}~\bibnamefont {Fujita}},
  \bibinfo {author} {\bibfnamefont {H.}~\bibnamefont {Yamaguchi}}, \bibinfo
  {author} {\bibfnamefont {D.}~\bibnamefont {Voiry}}, \bibinfo {author}
  {\bibfnamefont {M.}~\bibnamefont {Chen}}, \ and\ \bibinfo {author}
  {\bibfnamefont {M.}~\bibnamefont {Chhowalla}},\ }\href
  {http://dx.doi.org/10.1021/nn302422x} {\bibfield  {journal} {\bibinfo
  {journal} {ACS Nano}\ }\textbf {\bibinfo {volume} {6}},\ \bibinfo {pages}
  {7311} (\bibinfo {year} {2012})},\ \bibinfo {note} {pMID:
  22799455}\BibitemShut {NoStop}%
\bibitem [{\citenamefont {Guo}\ \emph {et~al.}(2015)\citenamefont {Guo},
  \citenamefont {Sun}, \citenamefont {Ouyang}, \citenamefont {Raja},
  \citenamefont {Song}, \citenamefont {Heinz},\ and\ \citenamefont
  {Brus}}]{Guo2015}%
  \BibitemOpen
  \bibfield  {author} {\bibinfo {author} {\bibfnamefont {Y.}~\bibnamefont
  {Guo}}, \bibinfo {author} {\bibfnamefont {D.}~\bibnamefont {Sun}}, \bibinfo
  {author} {\bibfnamefont {B.}~\bibnamefont {Ouyang}}, \bibinfo {author}
  {\bibfnamefont {A.}~\bibnamefont {Raja}}, \bibinfo {author} {\bibfnamefont
  {J.}~\bibnamefont {Song}}, \bibinfo {author} {\bibfnamefont {T.~F.}\
  \bibnamefont {Heinz}}, \ and\ \bibinfo {author} {\bibfnamefont {L.~E.}\
  \bibnamefont {Brus}},\ }\href {\doibase 10.1021/acs.nanolett.5b01196}
  {\bibfield  {journal} {\bibinfo  {journal} {Nano Lett.}\ }\textbf {\bibinfo
  {volume} {15}},\ \bibinfo {pages} {5081} (\bibinfo {year} {2015})},\ \bibinfo
  {note} {pMID: 26134736}\BibitemShut {NoStop}%
\bibitem [{\citenamefont {Gao}\ \emph {et~al.}(2015)\citenamefont {Gao},
  \citenamefont {Jiao}, \citenamefont {Ma}, \citenamefont {Jiao}, \citenamefont
  {Waclawik},\ and\ \citenamefont {Du}}]{gao2015charge}%
  \BibitemOpen
  \bibfield  {author} {\bibinfo {author} {\bibfnamefont {G.}~\bibnamefont
  {Gao}}, \bibinfo {author} {\bibfnamefont {Y.}~\bibnamefont {Jiao}}, \bibinfo
  {author} {\bibfnamefont {F.}~\bibnamefont {Ma}}, \bibinfo {author}
  {\bibfnamefont {Y.}~\bibnamefont {Jiao}}, \bibinfo {author} {\bibfnamefont
  {E.~R.}\ \bibnamefont {Waclawik}}, \ and\ \bibinfo {author} {\bibfnamefont
  {A.}~\bibnamefont {Du}},\ }\href
  {http://pubs.acs.org/doi/abs/10.1021/acs.jpcc.5b04658} {\bibfield  {journal}
  {\bibinfo  {journal} {J. Phys. Chem. C}\ }\textbf {\bibinfo {volume} {23}},\
  \bibinfo {pages} {13124} (\bibinfo {year} {2015})}\BibitemShut {NoStop}%
\bibitem [{\citenamefont {Kresse}\ and\ \citenamefont
  {Furthm\"uller}(1996{\natexlab{a}})}]{Kresse1996}%
  \BibitemOpen
  \bibfield  {author} {\bibinfo {author} {\bibfnamefont {G.}~\bibnamefont
  {Kresse}}\ and\ \bibinfo {author} {\bibfnamefont {J.}~\bibnamefont
  {Furthm\"uller}},\ }\href {\doibase 10.1103/PhysRevB.54.11169} {\bibfield
  {journal} {\bibinfo  {journal} {Phys. Rev. B}\ }\textbf {\bibinfo {volume}
  {54}},\ \bibinfo {pages} {11169} (\bibinfo {year}
  {1996}{\natexlab{a}})}\BibitemShut {NoStop}%
\bibitem [{\citenamefont {Kresse}\ and\ \citenamefont
  {Furthm\"uller}(1996{\natexlab{b}})}]{Kresse1996a}%
  \BibitemOpen
  \bibfield  {author} {\bibinfo {author} {\bibfnamefont {G.}~\bibnamefont
  {Kresse}}\ and\ \bibinfo {author} {\bibfnamefont {J.}~\bibnamefont
  {Furthm\"uller}},\ }\href {\doibase
  http://dx.doi.org/10.1016/0927-0256(96)00008-0} {\bibfield  {journal}
  {\bibinfo  {journal} {Comput. Phys. Sci.}\ }\textbf {\bibinfo {volume} {6}},\
  \bibinfo {pages} {15 } (\bibinfo {year} {1996}{\natexlab{b}})}\BibitemShut
  {NoStop}%
\bibitem [{\citenamefont {Bl\"ochl}(1994)}]{PAW}%
  \BibitemOpen
  \bibfield  {author} {\bibinfo {author} {\bibfnamefont {P.~E.}\ \bibnamefont
  {Bl\"ochl}},\ }\href
  {http://journals.aps.org/prb/abstract/10.1103/PhysRevB.50.17953} {\bibfield
  {journal} {\bibinfo  {journal} {Phys. Rev. B}\ }\textbf {\bibinfo {volume}
  {50}},\ \bibinfo {pages} {17953} (\bibinfo {year} {1994})}\BibitemShut
  {NoStop}%
\bibitem [{\citenamefont {Kresse}\ and\ \citenamefont
  {Joubert}(1999)}]{Kresse1999}%
  \BibitemOpen
  \bibfield  {author} {\bibinfo {author} {\bibfnamefont {G.}~\bibnamefont
  {Kresse}}\ and\ \bibinfo {author} {\bibfnamefont {D.}~\bibnamefont
  {Joubert}},\ }\href
  {http://journals.aps.org/prb/abstract/10.1103/PhysRevB.59.1758} {\bibfield
  {journal} {\bibinfo  {journal} {Phys. Rev. B}\ }\textbf {\bibinfo {volume}
  {59}},\ \bibinfo {pages} {1758} (\bibinfo {year} {1999})}\BibitemShut
  {NoStop}%
\bibitem [{\citenamefont {Perdew}, \citenamefont {Burke},\ and\ \citenamefont
  {Ernzerhof}(1996)}]{Perdew1996}%
  \BibitemOpen
  \bibfield  {author} {\bibinfo {author} {\bibfnamefont {J.~P.}\ \bibnamefont
  {Perdew}}, \bibinfo {author} {\bibfnamefont {K.}~\bibnamefont {Burke}}, \
  and\ \bibinfo {author} {\bibfnamefont {M.}~\bibnamefont {Ernzerhof}},\ }\href
  {http://journals.aps.org/prl/abstract/10.1103/PhysRevLett.77.3865} {\bibfield
   {journal} {\bibinfo  {journal} {Phys. Rev. Lett.}\ }\textbf {\bibinfo
  {volume} {77}},\ \bibinfo {pages} {3865} (\bibinfo {year}
  {1996})}\BibitemShut {NoStop}%
\bibitem [{\citenamefont {Heyd}, \citenamefont {Scuseria},\ and\ \citenamefont
  {Ernzerhof}(2003)}]{Heyd2003Hybrid}%
  \BibitemOpen
  \bibfield  {author} {\bibinfo {author} {\bibfnamefont {J.}~\bibnamefont
  {Heyd}}, \bibinfo {author} {\bibfnamefont {G.~E.}\ \bibnamefont {Scuseria}},
  \ and\ \bibinfo {author} {\bibfnamefont {M.}~\bibnamefont {Ernzerhof}},\
  }\href
  {http://scitation.aip.org/content/aip/journal/jcp/118/18/10.1063/1.1564060}
  {\bibfield  {journal} {\bibinfo  {journal} {J. Chem. Phys.}\ }\textbf
  {\bibinfo {volume} {118}},\ \bibinfo {pages} {8207} (\bibinfo {year}
  {2003})}\BibitemShut {NoStop}%
\bibitem [{\citenamefont {Heyd}, \citenamefont {Scuseria},\ and\ \citenamefont
  {Ernzerhof}(2006)}]{Heyd2006Erratum}%
  \BibitemOpen
  \bibfield  {author} {\bibinfo {author} {\bibfnamefont {J.}~\bibnamefont
  {Heyd}}, \bibinfo {author} {\bibfnamefont {G.~E.}\ \bibnamefont {Scuseria}},
  \ and\ \bibinfo {author} {\bibfnamefont {M.}~\bibnamefont {Ernzerhof}},\
  }\href
  {http://scitation.aip.org/content/aip/journal/jcp/124/21/10.1063/1.2204597}
  {\bibfield  {journal} {\bibinfo  {journal} {J. Chem. Phys.}\ }\textbf
  {\bibinfo {volume} {124}},\ \bibinfo {eid} {219906} (\bibinfo {year}
  {2006})}\BibitemShut {NoStop}%
\bibitem [{\citenamefont {Grimme}(2006)}]{Grimme2006}%
  \BibitemOpen
  \bibfield  {author} {\bibinfo {author} {\bibfnamefont {S.}~\bibnamefont
  {Grimme}},\ }\href {http://dx.doi.org/10.1002/jcc.20495} {\bibfield
  {journal} {\bibinfo  {journal} {J. Comput. Chem.}\ }\textbf {\bibinfo
  {volume} {27}},\ \bibinfo {pages} {1787} (\bibinfo {year}
  {2006})}\BibitemShut {NoStop}%
\bibitem [{\citenamefont {Bu{\v{c}}ko}\ \emph {et~al.}(2010)\citenamefont
  {Bu{\v{c}}ko}, \citenamefont {Hafner}, \citenamefont {Leb{\`e}gue},\ and\
  \citenamefont {{\'A}ngy{\'a}n}}]{Bucko2010}%
  \BibitemOpen
  \bibfield  {author} {\bibinfo {author} {\bibfnamefont {T.}~\bibnamefont
  {Bu{\v{c}}ko}}, \bibinfo {author} {\bibfnamefont {J.}~\bibnamefont {Hafner}},
  \bibinfo {author} {\bibfnamefont {S.}~\bibnamefont {Leb{\`e}gue}}, \ and\
  \bibinfo {author} {\bibfnamefont {J.}~\bibnamefont {{\'A}ngy{\'a}n}},\ }\href
  {\doibase 10.1021/jp106469x} {\bibfield  {journal} {\bibinfo  {journal} {J.
  Phys. Chem. A}\ }\textbf {\bibinfo {volume} {114}},\ \bibinfo {pages} {11814}
  (\bibinfo {year} {2010})}\BibitemShut {NoStop}%
\bibitem [{\citenamefont {Monkhorst}\ and\ \citenamefont
  {Pack}(1976)}]{Monkhorst1976}%
  \BibitemOpen
  \bibfield  {author} {\bibinfo {author} {\bibfnamefont {H.~J.}\ \bibnamefont
  {Monkhorst}}\ and\ \bibinfo {author} {\bibfnamefont {J.~D.}\ \bibnamefont
  {Pack}},\ }\href {\doibase 10.1103/PhysRevB.13.5188} {\bibfield  {journal}
  {\bibinfo  {journal} {Phys. Rev. B}\ }\textbf {\bibinfo {volume} {13}},\
  \bibinfo {pages} {5188} (\bibinfo {year} {1976})}\BibitemShut {NoStop}%
\bibitem [{\citenamefont {Togo}\ and\ \citenamefont {Tanaka}(2015)}]{phonopy}%
  \BibitemOpen
  \bibfield  {author} {\bibinfo {author} {\bibfnamefont {A.}~\bibnamefont
  {Togo}}\ and\ \bibinfo {author} {\bibfnamefont {I.}~\bibnamefont {Tanaka}},\
  }\href {http://www.sciencedirect.com/science/article/pii/S1359646215003127}
  {\bibfield  {journal} {\bibinfo  {journal} {Scr. Mater.}\ }\textbf {\bibinfo
  {volume} {108}},\ \bibinfo {pages} {1} (\bibinfo {year} {2015})}\BibitemShut
  {NoStop}%
\bibitem [{\citenamefont {Gonze}\ and\ \citenamefont {Lee}(1997)}]{Gonze1997}%
  \BibitemOpen
  \bibfield  {author} {\bibinfo {author} {\bibfnamefont {X.}~\bibnamefont
  {Gonze}}\ and\ \bibinfo {author} {\bibfnamefont {C.}~\bibnamefont {Lee}},\
  }\href {\doibase 10.1103/PhysRevB.55.10355} {\bibfield  {journal} {\bibinfo
  {journal} {Phys. Rev. B}\ }\textbf {\bibinfo {volume} {55}},\ \bibinfo
  {pages} {10355} (\bibinfo {year} {1997})}\BibitemShut {NoStop}%
\bibitem [{\citenamefont {Tom\'anek}\ and\ \citenamefont
  {Louie}(1988)}]{Tomanek1988}%
  \BibitemOpen
  \bibfield  {author} {\bibinfo {author} {\bibfnamefont {D.}~\bibnamefont
  {Tom\'anek}}\ and\ \bibinfo {author} {\bibfnamefont {S.~G.}\ \bibnamefont
  {Louie}},\ }\href {\doibase 10.1103/PhysRevB.37.8327} {\bibfield  {journal}
  {\bibinfo  {journal} {Phys. Rev. B}\ }\textbf {\bibinfo {volume} {37}},\
  \bibinfo {pages} {8327} (\bibinfo {year} {1988})}\BibitemShut {NoStop}%
\bibitem [{\citenamefont {Selloni}\ \emph {et~al.}(1985)\citenamefont
  {Selloni}, \citenamefont {Carnevali}, \citenamefont {Tosatti},\ and\
  \citenamefont {Chen}}]{Selloni1985}%
  \BibitemOpen
  \bibfield  {author} {\bibinfo {author} {\bibfnamefont {A.}~\bibnamefont
  {Selloni}}, \bibinfo {author} {\bibfnamefont {P.}~\bibnamefont {Carnevali}},
  \bibinfo {author} {\bibfnamefont {E.}~\bibnamefont {Tosatti}}, \ and\
  \bibinfo {author} {\bibfnamefont {C.~D.}\ \bibnamefont {Chen}},\ }\href
  {\doibase 10.1103/PhysRevB.31.2602} {\bibfield  {journal} {\bibinfo
  {journal} {Phys. Rev. B}\ }\textbf {\bibinfo {volume} {31}},\ \bibinfo
  {pages} {2602} (\bibinfo {year} {1985})}\BibitemShut {NoStop}%
\bibitem [{\citenamefont {Tersoff}\ and\ \citenamefont
  {Hamann}(1983)}]{Tersoff1983}%
  \BibitemOpen
  \bibfield  {author} {\bibinfo {author} {\bibfnamefont {J.}~\bibnamefont
  {Tersoff}}\ and\ \bibinfo {author} {\bibfnamefont {D.~R.}\ \bibnamefont
  {Hamann}},\ }\href {\doibase 10.1103/PhysRevLett.50.1998} {\bibfield
  {journal} {\bibinfo  {journal} {Phys. Rev. Lett.}\ }\textbf {\bibinfo
  {volume} {50}},\ \bibinfo {pages} {1998} (\bibinfo {year}
  {1983})}\BibitemShut {NoStop}%
\bibitem [{\citenamefont {Tersoff}\ and\ \citenamefont
  {Hamann}(1985)}]{Tersoff1985}%
  \BibitemOpen
  \bibfield  {author} {\bibinfo {author} {\bibfnamefont {J.}~\bibnamefont
  {Tersoff}}\ and\ \bibinfo {author} {\bibfnamefont {D.~R.}\ \bibnamefont
  {Hamann}},\ }\href {\doibase 10.1103/PhysRevB.31.805} {\bibfield  {journal}
  {\bibinfo  {journal} {Phys. Rev. B}\ }\textbf {\bibinfo {volume} {31}},\
  \bibinfo {pages} {805} (\bibinfo {year} {1985})}\BibitemShut {NoStop}%
\bibitem [{\citenamefont {Fuhr}, \citenamefont {Sa\'ul},\ and\ \citenamefont
  {Sofo}(2004)}]{Fuhr2004}%
  \BibitemOpen
  \bibfield  {author} {\bibinfo {author} {\bibfnamefont {J.~D.}\ \bibnamefont
  {Fuhr}}, \bibinfo {author} {\bibfnamefont {A.}~\bibnamefont {Sa\'ul}}, \ and\
  \bibinfo {author} {\bibfnamefont {J.~O.}\ \bibnamefont {Sofo}},\ }\href
  {\doibase 10.1103/PhysRevLett.92.026802} {\bibfield  {journal} {\bibinfo
  {journal} {Phys. Rev. Lett.}\ }\textbf {\bibinfo {volume} {92}},\ \bibinfo
  {pages} {026802} (\bibinfo {year} {2004})}\BibitemShut {NoStop}%
\bibitem [{\citenamefont {Rivero}\ \emph {et~al.}(2015)\citenamefont {Rivero},
  \citenamefont {Horvath}, \citenamefont {Zhu}, \citenamefont {Guan},
  \citenamefont {Tom\'anek},\ and\ \citenamefont {Barraza-Lopez}}]{Rivero2015}%
  \BibitemOpen
  \bibfield  {author} {\bibinfo {author} {\bibfnamefont {P.}~\bibnamefont
  {Rivero}}, \bibinfo {author} {\bibfnamefont {C.~M.}\ \bibnamefont {Horvath}},
  \bibinfo {author} {\bibfnamefont {Z.}~\bibnamefont {Zhu}}, \bibinfo {author}
  {\bibfnamefont {J.}~\bibnamefont {Guan}}, \bibinfo {author} {\bibfnamefont
  {D.}~\bibnamefont {Tom\'anek}}, \ and\ \bibinfo {author} {\bibfnamefont
  {S.}~\bibnamefont {Barraza-Lopez}},\ }\href {\doibase
  10.1103/PhysRevB.91.115413} {\bibfield  {journal} {\bibinfo  {journal} {Phys.
  Rev. B}\ }\textbf {\bibinfo {volume} {91}},\ \bibinfo {pages} {115413}
  (\bibinfo {year} {2015})}\BibitemShut {NoStop}%
\bibitem [{\citenamefont {Chhowalla}\ \emph {et~al.}(2013)\citenamefont
  {Chhowalla}, \citenamefont {Shin}, \citenamefont {Eda}, \citenamefont {Li},
  \citenamefont {Loh},\ and\ \citenamefont {Zhang}}]{chhowalla2013}%
  \BibitemOpen
  \bibfield  {author} {\bibinfo {author} {\bibfnamefont {M.}~\bibnamefont
  {Chhowalla}}, \bibinfo {author} {\bibfnamefont {H.~S.}\ \bibnamefont {Shin}},
  \bibinfo {author} {\bibfnamefont {G.}~\bibnamefont {Eda}}, \bibinfo {author}
  {\bibfnamefont {L.-J.}\ \bibnamefont {Li}}, \bibinfo {author} {\bibfnamefont
  {K.~P.}\ \bibnamefont {Loh}}, \ and\ \bibinfo {author} {\bibfnamefont
  {H.}~\bibnamefont {Zhang}},\ }\href
  {http://www.nature.com/nchem/journal/v5/n4/full/nchem.1589.html} {\bibfield
  {journal} {\bibinfo  {journal} {Nat. Chem.}\ }\textbf {\bibinfo {volume}
  {5}},\ \bibinfo {pages} {263} (\bibinfo {year} {2013})}\BibitemShut {NoStop}%
\bibitem [{\citenamefont {Mahler}\ \emph {et~al.}(2014)\citenamefont {Mahler},
  \citenamefont {Hoepfner}, \citenamefont {Liao},\ and\ \citenamefont
  {Ozin}}]{Mahler2014}%
  \BibitemOpen
  \bibfield  {author} {\bibinfo {author} {\bibfnamefont {B.}~\bibnamefont
  {Mahler}}, \bibinfo {author} {\bibfnamefont {V.}~\bibnamefont {Hoepfner}},
  \bibinfo {author} {\bibfnamefont {K.}~\bibnamefont {Liao}}, \ and\ \bibinfo
  {author} {\bibfnamefont {G.~A.}\ \bibnamefont {Ozin}},\ }\href
  {http://dx.doi.org/10.1021/ja506261t} {\bibfield  {journal} {\bibinfo
  {journal} {J. Am. Chem. Soc.}\ }\textbf {\bibinfo {volume} {136}},\ \bibinfo
  {pages} {14121} (\bibinfo {year} {2014})},\ \bibinfo {note} {pMID:
  25220034}\BibitemShut {NoStop}%
\bibitem [{\citenamefont {Keum}\ \emph {et~al.}(2015)\citenamefont {Keum},
  \citenamefont {Cho}, \citenamefont {Kim}, \citenamefont {Choe}, \citenamefont
  {Sung}, \citenamefont {Kan}, \citenamefont {Kang}, \citenamefont {Hwang},
  \citenamefont {Kim}, \citenamefont {Yang},\ and\ \citenamefont
  {et~al.}}]{Keum2015}%
  \BibitemOpen
  \bibfield  {author} {\bibinfo {author} {\bibfnamefont {D.~H.}\ \bibnamefont
  {Keum}}, \bibinfo {author} {\bibfnamefont {S.}~\bibnamefont {Cho}}, \bibinfo
  {author} {\bibfnamefont {J.~H.}\ \bibnamefont {Kim}}, \bibinfo {author}
  {\bibfnamefont {D.-H.}\ \bibnamefont {Choe}}, \bibinfo {author}
  {\bibfnamefont {H.-J.}\ \bibnamefont {Sung}}, \bibinfo {author}
  {\bibfnamefont {M.}~\bibnamefont {Kan}}, \bibinfo {author} {\bibfnamefont
  {H.}~\bibnamefont {Kang}}, \bibinfo {author} {\bibfnamefont {J.-Y.}\
  \bibnamefont {Hwang}}, \bibinfo {author} {\bibfnamefont {S.~W.}\ \bibnamefont
  {Kim}}, \bibinfo {author} {\bibfnamefont {H.}~\bibnamefont {Yang}}, \ and\
  \bibinfo {author} {\bibnamefont {et~al.}},\ }\href {\doibase
  10.1038/nphys3314} {\bibfield  {journal} {\bibinfo  {journal} {Nat. Phys.}\
  }\textbf {\bibinfo {volume} {11}},\ \bibinfo {pages} {482} (\bibinfo {year}
  {2015})}\BibitemShut {NoStop}%
\bibitem [{\citenamefont {Altibelli}, \citenamefont {Joachim},\ and\
  \citenamefont {Sautet}(1996)}]{Altibelli1996}%
  \BibitemOpen
  \bibfield  {author} {\bibinfo {author} {\bibfnamefont {A.}~\bibnamefont
  {Altibelli}}, \bibinfo {author} {\bibfnamefont {C.}~\bibnamefont {Joachim}},
  \ and\ \bibinfo {author} {\bibfnamefont {P.}~\bibnamefont {Sautet}},\ }\href
  {\doibase http://dx.doi.org/10.1016/S0039-6028(96)00864-3} {\bibfield
  {journal} {\bibinfo  {journal} {Surf. Sci.}\ }\textbf {\bibinfo {volume}
  {367}},\ \bibinfo {pages} {209 } (\bibinfo {year} {1996})}\BibitemShut
  {NoStop}%
\bibitem [{\citenamefont {Yang}\ \emph {et~al.}(1991)\citenamefont {Yang},
  \citenamefont {Sandoval}, \citenamefont {Divigalpitiya}, \citenamefont
  {Irwin},\ and\ \citenamefont {Frindt}}]{Yang1991}%
  \BibitemOpen
  \bibfield  {author} {\bibinfo {author} {\bibfnamefont {D.}~\bibnamefont
  {Yang}}, \bibinfo {author} {\bibfnamefont {S.~J.}\ \bibnamefont {Sandoval}},
  \bibinfo {author} {\bibfnamefont {W.~M.~R.}\ \bibnamefont {Divigalpitiya}},
  \bibinfo {author} {\bibfnamefont {J.~C.}\ \bibnamefont {Irwin}}, \ and\
  \bibinfo {author} {\bibfnamefont {R.~F.}\ \bibnamefont {Frindt}},\ }\href
  {\doibase 10.1103/PhysRevB.43.12053} {\bibfield  {journal} {\bibinfo
  {journal} {Phys. Rev. B}\ }\textbf {\bibinfo {volume} {43}},\ \bibinfo
  {pages} {12053} (\bibinfo {year} {1991})}\BibitemShut {NoStop}%
\bibitem [{\citenamefont {Gordon}\ \emph {et~al.}(2002)\citenamefont {Gordon},
  \citenamefont {Yang}, \citenamefont {Crozier}, \citenamefont {Jiang},\ and\
  \citenamefont {Frindt}}]{Gordon2002}%
  \BibitemOpen
  \bibfield  {author} {\bibinfo {author} {\bibfnamefont {R.~A.}\ \bibnamefont
  {Gordon}}, \bibinfo {author} {\bibfnamefont {D.}~\bibnamefont {Yang}},
  \bibinfo {author} {\bibfnamefont {E.~D.}\ \bibnamefont {Crozier}}, \bibinfo
  {author} {\bibfnamefont {D.~T.}\ \bibnamefont {Jiang}}, \ and\ \bibinfo
  {author} {\bibfnamefont {R.~F.}\ \bibnamefont {Frindt}},\ }\href {\doibase
  10.1103/PhysRevB.65.125407} {\bibfield  {journal} {\bibinfo  {journal} {Phys.
  Rev. B}\ }\textbf {\bibinfo {volume} {65}},\ \bibinfo {pages} {125407}
  (\bibinfo {year} {2002})}\BibitemShut {NoStop}%
\bibitem [{\citenamefont {Ding}\ and\ \citenamefont {Wang}(2013)}]{Ding2013}%
  \BibitemOpen
  \bibfield  {author} {\bibinfo {author} {\bibfnamefont {Y.}~\bibnamefont
  {Ding}}\ and\ \bibinfo {author} {\bibfnamefont {Y.}~\bibnamefont {Wang}},\
  }\href {\doibase 10.1021/jp407666m} {\bibfield  {journal} {\bibinfo
  {journal} {J. Phys. Chem. C}\ }\textbf {\bibinfo {volume} {117}},\ \bibinfo
  {pages} {18266} (\bibinfo {year} {2013})}\BibitemShut {NoStop}%
\bibitem [{\citenamefont {Wang}\ \emph
  {et~al.}(2015{\natexlab{a}})\citenamefont {Wang}, \citenamefont {Kutana},
  \citenamefont {Zou},\ and\ \citenamefont {Yakobson}}]{Wang2015}%
  \BibitemOpen
  \bibfield  {author} {\bibinfo {author} {\bibfnamefont {L.}~\bibnamefont
  {Wang}}, \bibinfo {author} {\bibfnamefont {A.}~\bibnamefont {Kutana}},
  \bibinfo {author} {\bibfnamefont {X.}~\bibnamefont {Zou}}, \ and\ \bibinfo
  {author} {\bibfnamefont {B.~I.}\ \bibnamefont {Yakobson}},\ }\href {\doibase
  10.1039/C5NR00355E} {\bibfield  {journal} {\bibinfo  {journal} {Nanoscale}\
  }\textbf {\bibinfo {volume} {7}},\ \bibinfo {pages} {9746} (\bibinfo {year}
  {2015}{\natexlab{a}})}\BibitemShut {NoStop}%
\bibitem [{\citenamefont {Andrew}\ \emph {et~al.}(2012)\citenamefont {Andrew},
  \citenamefont {Mapasha}, \citenamefont {Ukpong},\ and\ \citenamefont
  {Chetty}}]{PhysRevB.85.125428}%
  \BibitemOpen
  \bibfield  {author} {\bibinfo {author} {\bibfnamefont {R.~C.}\ \bibnamefont
  {Andrew}}, \bibinfo {author} {\bibfnamefont {R.~E.}\ \bibnamefont {Mapasha}},
  \bibinfo {author} {\bibfnamefont {A.~M.}\ \bibnamefont {Ukpong}}, \ and\
  \bibinfo {author} {\bibfnamefont {N.}~\bibnamefont {Chetty}},\ }\href
  {\doibase 10.1103/PhysRevB.85.125428} {\bibfield  {journal} {\bibinfo
  {journal} {Phys. Rev. B}\ }\textbf {\bibinfo {volume} {85}},\ \bibinfo
  {pages} {125428} (\bibinfo {year} {2012})}\BibitemShut {NoStop}%
\bibitem [{\citenamefont {Wang}\ and\ \citenamefont
  {Geng}(2016)}]{wang2016lattice}%
  \BibitemOpen
  \bibfield  {author} {\bibinfo {author} {\bibfnamefont {V.}~\bibnamefont
  {Wang}}\ and\ \bibinfo {author} {\bibfnamefont {W.}~\bibnamefont {Geng}},\
  }\href {http://arxiv.org/abs/1607.00642} {\bibfield  {journal} {\bibinfo
  {journal} {arXiv preprint arXiv:1607.00642}\ } (\bibinfo {year}
  {2016})}\BibitemShut {NoStop}%
\bibitem [{\citenamefont {Jim\'enez~Sandoval}\ \emph
  {et~al.}(1991)\citenamefont {Jim\'enez~Sandoval}, \citenamefont {Yang},
  \citenamefont {Frindt},\ and\ \citenamefont {Irwin}}]{JimenezSandoval1991}%
  \BibitemOpen
  \bibfield  {author} {\bibinfo {author} {\bibfnamefont {S.}~\bibnamefont
  {Jim\'enez~Sandoval}}, \bibinfo {author} {\bibfnamefont {D.}~\bibnamefont
  {Yang}}, \bibinfo {author} {\bibfnamefont {R.~F.}\ \bibnamefont {Frindt}}, \
  and\ \bibinfo {author} {\bibfnamefont {J.~C.}\ \bibnamefont {Irwin}},\ }\href
  {\doibase 10.1103/PhysRevB.44.3955} {\bibfield  {journal} {\bibinfo
  {journal} {Phys. Rev. B}\ }\textbf {\bibinfo {volume} {44}},\ \bibinfo
  {pages} {3955} (\bibinfo {year} {1991})}\BibitemShut {NoStop}%
\bibitem [{\citenamefont {Molina-S\'anchez}\ and\ \citenamefont
  {Wirtz}(2011)}]{Molina-Sanchez2011}%
  \BibitemOpen
  \bibfield  {author} {\bibinfo {author} {\bibfnamefont {A.}~\bibnamefont
  {Molina-S\'anchez}}\ and\ \bibinfo {author} {\bibfnamefont {L.}~\bibnamefont
  {Wirtz}},\ }\href {\doibase 10.1103/PhysRevB.84.155413} {\bibfield  {journal}
  {\bibinfo  {journal} {Phys. Rev. B}\ }\textbf {\bibinfo {volume} {84}},\
  \bibinfo {pages} {155413} (\bibinfo {year} {2011})}\BibitemShut {NoStop}%
\bibitem [{\citenamefont {Dresselhaus}, \citenamefont {Dresselhaus},\ and\
  \citenamefont {Jorio}(2007)}]{Dresselhaus2007group}%
  \BibitemOpen
  \bibfield  {author} {\bibinfo {author} {\bibfnamefont {M.~S.}\ \bibnamefont
  {Dresselhaus}}, \bibinfo {author} {\bibfnamefont {G.}~\bibnamefont
  {Dresselhaus}}, \ and\ \bibinfo {author} {\bibfnamefont {A.}~\bibnamefont
  {Jorio}},\ }\href {http://adsabs.harvard.edu/abs/2008PhT....61k..57D} {\emph
  {\bibinfo {title} {Group theory: application to the physics of condensed
  matter}}}\ (\bibinfo  {publisher} {Springer Science \& Business Media},\
  \bibinfo {year} {2007})\BibitemShut {NoStop}%
\bibitem [{\citenamefont {Wieting}\ and\ \citenamefont
  {Verble}(1971)}]{Wieting1971}%
  \BibitemOpen
  \bibfield  {author} {\bibinfo {author} {\bibfnamefont {T.~J.}\ \bibnamefont
  {Wieting}}\ and\ \bibinfo {author} {\bibfnamefont {J.~L.}\ \bibnamefont
  {Verble}},\ }\href {\doibase 10.1103/PhysRevB.3.4286} {\bibfield  {journal}
  {\bibinfo  {journal} {Phys. Rev. B}\ }\textbf {\bibinfo {volume} {3}},\
  \bibinfo {pages} {4286} (\bibinfo {year} {1971})}\BibitemShut {NoStop}%
\bibitem [{\citenamefont {Cai}\ \emph {et~al.}(2014)\citenamefont {Cai},
  \citenamefont {Lan}, \citenamefont {Zhang},\ and\ \citenamefont
  {Zhang}}]{Cai2014}%
  \BibitemOpen
  \bibfield  {author} {\bibinfo {author} {\bibfnamefont {Y.}~\bibnamefont
  {Cai}}, \bibinfo {author} {\bibfnamefont {J.}~\bibnamefont {Lan}}, \bibinfo
  {author} {\bibfnamefont {G.}~\bibnamefont {Zhang}}, \ and\ \bibinfo {author}
  {\bibfnamefont {Y.-W.}\ \bibnamefont {Zhang}},\ }\href {\doibase
  10.1103/PhysRevB.89.035438} {\bibfield  {journal} {\bibinfo  {journal} {Phys.
  Rev. B}\ }\textbf {\bibinfo {volume} {89}},\ \bibinfo {pages} {035438}
  (\bibinfo {year} {2014})}\BibitemShut {NoStop}%
\bibitem [{\citenamefont {Zhang}\ \emph {et~al.}(2015)\citenamefont {Zhang},
  \citenamefont {Qiao}, \citenamefont {Shi}, \citenamefont {Wu}, \citenamefont
  {Jiang},\ and\ \citenamefont {Tan}}]{Zhang2015}%
  \BibitemOpen
  \bibfield  {author} {\bibinfo {author} {\bibfnamefont {X.}~\bibnamefont
  {Zhang}}, \bibinfo {author} {\bibfnamefont {X.-F.}\ \bibnamefont {Qiao}},
  \bibinfo {author} {\bibfnamefont {W.}~\bibnamefont {Shi}}, \bibinfo {author}
  {\bibfnamefont {J.-B.}\ \bibnamefont {Wu}}, \bibinfo {author} {\bibfnamefont
  {D.-S.}\ \bibnamefont {Jiang}}, \ and\ \bibinfo {author} {\bibfnamefont
  {P.-H.}\ \bibnamefont {Tan}},\ }\href {\doibase 10.1039/C4CS00282B}
  {\bibfield  {journal} {\bibinfo  {journal} {Chem. Soc. Rev.}\ }\textbf
  {\bibinfo {volume} {44}},\ \bibinfo {pages} {2757} (\bibinfo {year}
  {2015})}\BibitemShut {NoStop}%
\bibitem [{\citenamefont {Wang}\ \emph
  {et~al.}(2015{\natexlab{b}})\citenamefont {Wang}, \citenamefont {Liu},
  \citenamefont {Kawazoe},\ and\ \citenamefont {Geng}}]{Wang2015Role}%
  \BibitemOpen
  \bibfield  {author} {\bibinfo {author} {\bibfnamefont {V.}~\bibnamefont
  {Wang}}, \bibinfo {author} {\bibfnamefont {Y.~C.}\ \bibnamefont {Liu}},
  \bibinfo {author} {\bibfnamefont {Y.}~\bibnamefont {Kawazoe}}, \ and\
  \bibinfo {author} {\bibfnamefont {W.~T.}\ \bibnamefont {Geng}},\ }\href
  {\doibase 10.1021/acs.jpclett.5b02047} {\bibfield  {journal} {\bibinfo
  {journal} {J. Phys. Chem. Lett.}\ }\textbf {\bibinfo {volume} {6}},\ \bibinfo
  {pages} {4876} (\bibinfo {year} {2015}{\natexlab{b}})}\BibitemShut {NoStop}%
\end{thebibliography}%
\end{document}